\documentclass[12pt]{article}

\pdfoutput=1

\usepackage{graphics}
\usepackage{epsfig}

\usepackage{amsfonts}
\usepackage{amssymb}
\usepackage{amsmath}
\usepackage{dsfont}
\usepackage{pifont}
\usepackage{bbm}
\usepackage{multirow}
\usepackage{latexsym}
\usepackage{verbatim}
\usepackage{mcite}
\usepackage{cite}
\usepackage{units}
\usepackage[all]{xy}
\usepackage{cancel}
\usepackage{slashed}
\usepackage{textcomp}

\def\beq{\begin{equation}}
\def\eeq{\end{equation}}
\def\beqa{\begin{eqnarray}}
\def\eeqa{\end{eqnarray}}

\def\a{{\alpha}}

\def\g{{\gamma}}

\def\d{{\delta}}

\def\m{{\mu}}
\def\n{{\nu}}
\def\r{{\rho}}
\def\s{{\sigma}}

\def\bfone{\relax{\rm 1\kern-.35em 1}}

\newcommand{\be}{\begin{equation}}
\newcommand{\ee}{\end{equation}}
\newcommand{\ben}{\begin{displaymath}}
\newcommand{\een}{\end{displaymath}}
\newcommand{\bea}{\begin{eqnarray}}
\newcommand{\eea}{\end{eqnarray}}

\newcommand{\bean}{\begin{eqnarray*}}
\newcommand{\eean}{\end{eqnarray*}}

\DeclareMathAlphabet{\mathpzc}{OT1}{pzc}{m}{it}

\begin{document}
\pagestyle{plain}

%----------------------------------------------------------------------%
%  numbering sections, equations, footnotes, etc...
%----------------------------------------------------------------------%

\makeatletter \@addtoreset{equation}{section} \makeatother
\renewcommand{\thesection}{\arabic{section}}
\renewcommand{\theequation}{\thesection.\arabic{equation}}
\renewcommand{\thefootnote}{\arabic{footnote}}

%----------------------------------------------------------------------%
%  Resetting of counters
%----------------------------------------------------------------------%

\setcounter{page}{1} \setcounter{footnote}{0}

%----------------------------------------------------------------------%
%  title page
%----------------------------------------------------------------------%

\begin{titlepage}
\begin{flushright}
UUITP-22/17\\
\end{flushright}

\bigskip

\begin{center}

\vskip 0cm

{\LARGE \bf  BPS objects in $D=7$ supergravity \\ and \\[2mm] their M-theory origin} \\[6mm]

\vskip 0.5cm

{\bf Giuseppe Dibitetto$^1$ \,and\, Nicol\`o Petri$^2$}\let\thefootnote\relax\footnote{giuseppe.dibitetto@physics.uu.se, nicolo.petri@mi.infn.it}\\

\vskip 0.5cm

{\em 
$^1$Institutionen f\"or fysik och astronomi, University of Uppsala, \\ Box 803, SE-751 08 Uppsala, Sweden \\
$^2$Dipartimento di Fisica, Universit\`a di Milano, and INFN, Sezione di Milano,\\ Via Celoria 16, I-20133 Milano, Italy}

\vskip 0.8cm

\end{center}

\vskip 1cm

\begin{center}

{\bf ABSTRACT}\\[3ex]

\begin{minipage}{13cm}
\small

We study several different types of BPS flows within minimal $\mathcal{N}=1$, $D=7$ supergravity with $\textrm{SU}(2)$ gauge group and non-vanishing topological mass.
After reviewing some known domain wall solutions involving only the metric and the $\mathbb{R}^{+}$ scalar field, we move to considering more general flows involving a ``dyonic'' profile for the 
3-form gauge potential. In this context, we consider flows featuring a $\textrm{Mkw}_{3}$ as well as an $\textrm{AdS}_{3}$ slicing, write down the corresponding flow equations, and integrate them 
analytically to obtain many examples of asymptotically $\mathrm{AdS}_7$ solutions in presence of a running 3-form. Furthermore, we move to adding the possibility of non-vanishing vector fields, find the new corresponding flows and integrate them numerically.
Finally, we discuss the eleven-dimensional interpretation of the aforementioned solutions as effective descriptions of $\mathrm{M2}-\mathrm{M5}$ bound states.

\end{minipage}

\end{center}

\vfill

\end{titlepage}

%%%%%%%%%%%%%%%%%%%%%%%%%%%%%%%%%%%%%%%%%%%%%%%%%%%%%%%%%
%%
%%               Contents
%%
%%%%%%%%%%%%%%%%%%%%%%%%%%%%%%%%%%%%%%%%%%%%%%%%%%%%%%%%%

\tableofcontents

%%%%%%%%%%%%%%%%%%%%%%%%%%%%%%%%%%%%%%%%%%%%%%%%%%%%%%%%%
%%
%%               Main body
%%
%%%%%%%%%%%%%%%%%%%%%%%%%%%%%%%%%%%%%%%%%%%%%%%%%%%%%%%%%

\section{Introduction}
\label{sec:introduction}

Gauged supergravities in dimensions lower than ten represent an extremely valuable tool for studying the properties of all flux compactifications in string theory that preserve some residual supersymmetry.
These theories may be regarded as deformed versions of supergravity where the gauging and all other consistent massive deformations encode all information concerning a given flux background, as well as
the geometry and topology of the compact manifold. 

Among the best understood and most celebrated examples of the above claim, we may certainly name the consistent sphere reductions of type II and 11D supergravity yielding maximally supersymmetric AdS 
vacua which are at the core of the AdS/CFT correspondence \cite{Maldacena:1997re}. In particular here, we refer to type IIB supergravity on $S^{5}$ \cite{Cvetic:2000nc}, and to 11D supergravity on 
$S^{4}$ \cite{Pilch:1984xy,Nastase:1999cb} and $S^{7}$ \cite{deWit:1986oxb}. All of these compactifications admit a gauged supergravity as lower-dimensional effective description where the gauge group is 
$\textrm{SO}(d+1)\,\equiv\,\textrm{ISO}(S^{d})$.
Moving to cases with less than maximal supersymmetry, we find consistent reductions on squashed spheres such as \emph{e.g.} massive type IIA on $S^{4}$ \cite{Brandhuber:1999np} and $S^{6}$ 
\cite{Guarino:2015vca}.

While all of the above compactifications are described by gauged maximal supergravities and truncations thereof, there exist many other classes which involve \emph{explicit} supersymmetry breaking due to
the presence of spacetime-filling branes and/or O-planes. Such situations admit gauged supergravities with a lower amount of supersymmetries as effective descriptions. Typical examples in this class can
be twisted tori \cite{Scherk:1979zr}, with extra $p$-form gauge fluxes. These reductions have been a very succcessful playground for moduli stabilization \cite{Villadoro:2005cu} due to their simplicity, 
since their consistency immediately follows from group-theoretical arguments. In general, as opposed to sphere reductions though, such compactifications yield non-semisimple gauge groups.

An important milestone in our way of describing, classifying and analyzing gauged supergravities is represented by the work of \cite{deWit:2002vt,deWit:2005ub}, which gave birth to the so-called 
\emph{embedding tensor formalism} as a way of comprizing all possible consistent gaugings of a supergravity theory under a unique universal formulation. The idea is based on restoring the full global 
symmetry of the deformed Lagrangian by promoting the deformation parameters to tensors w.r.t. the duality symmetry of the theory, where the term ``duality'' here is intentionally used to remind the 
reader that string dualities are realized as actual symmetries upon compactification.

This naturally results in a precise correspondence between gaugings and generalized fluxes, which is corroborated by the existence of a consistent group-theoretical prescription for deriving the 
embedding tensor/ fluxes dictionary (see \emph{e.g.} \cite{Samtleben:2008pe} for a nice review). In the particular context of massive type IIA compactifications many things have been worked out in detail
and can be found in \cite{DallAgata:2009wsi,Dibitetto:2010rg,Dibitetto:2011gm,Dibitetto:2014sfa}.

Focusing in particular on gauged supergravities in seven dimensions, they may be divided into maximal theories, \emph{i.e.} with $32$ real supercharges, and half-maximal ones with only $16$. Since Majorana
spinors do not exist in $1+6$ dimensions, it is impossible to further go down to $8$ supersymmetries. 
While the complete embedding tensor formulation of the maximal gauged theories has been worked out in all details in \cite{Samtleben:2005bp}, such a complete formulation is lacking in the context of 
theories with $16$ supercharges. However, some salient features were presented in \cite{Dibitetto:2012rk,Dibitetto:2015bia}, including a study of vacua.

The theories of interest in this paper will be particular truncations of half-maximal supergravities obtained by restricting oneself to the $\mathcal{N}=1$ supergravity multiplet. The theory in its 
ungauged incarnation has a bosonic field content comprizing the metric, a three-form gauge potential, three vector fields and one scalar, and is usually referred to as \emph{minimal}. 
The most general consistent deformation turns out to be a combination of a gauging of the $\textrm{SU}(2)$ R-symmetry group and a St\"uckelberg-like massive deformation for the three-form potential.
The purely gauged minimal theory was found to stem from a reduction of type I supergravity on $S^{3}$ \cite{Chamseddine:1999uy}, while the purely massive theory may be obtained as a reduction of 
eleven-dimensional supergravity on a $\mathbb{T}^{4}$ with non-vanishing four-form flux. 
However, none of the above limiting cases allows for moduli stabilization, since the induced scalar potential always exhibits a run-away direction.

Conversely, when turning on both deformations at the same time, the scalar potential possesses two AdS extrema, one of which is supersymmetric, the other one having spontaneously broken supersymmetry 
\cite{Mezincescu:1984ta}. This particular theory, besides admitting an uplift to eleven-dimensional supergravity on a squashed $S^{4}$ \cite{Lu:1999bc}, it was furthermore recently found to admit a
ten-dimensional origin from massive type IIA supergravity on a squashed $S^{3}$ \cite{Passias:2015gya} linking these AdS$_7$ solutions to those in \cite{Apruzzi:2013yva}. 

When moving away from the study of vacua to more general BPS flows, the simplest type of solutions which one encounters are domain walls (DW), where the metric and the scalar fields assume a
non-trivial profile. In the context of maximal $D=7$ supergravity, the DW solutions for all $\textrm{CSO}(p,q,5-p-q)$-gauged theories were classified and given a higher-dimensional origin as branes 
reduced on their transverse space \cite{Bergshoeff:2004nq}.

Ever since the work of \cite{Maldacena:2000mw}, more general flows involving vector fields were found, describing spontaneous compactifications of $\textrm{AdS}_{7}$ down to lower-dimensional AdS spaces. 
More examples in this class were found in \cite{Gauntlett:2006ux,Bah:2011vv,Bah:2012dg,Bah:2013aha}, where the solutions are furthermore physically interpreted as IR conformal fixed points obtained via
M5-brane wrapping. 
More recently in \cite{Karndumri:2014hma,Karndumri:2015rsa} and \cite{Rota:2015aoa,Passias:2015gya,Bah:2017wxp}, analogous BPS flows were presented within half-maximal gauged supergravity, respectively 
coupled to vector multiplets and minimal.

The goal of our work is that of extending the above classes of BPS flows in $\mathcal{N}=1$, $D=7$ gauged supergravities by including novel examples with a non-trivial profile for the three-form gauge 
potential. To this end, we will first review some known BPS DW solutions and then move to more general flows involving the three-form, both with and without vector fields turned on. 
Most of the flow equations that we write down will be then integrated analytically, while for some of them we will have to employ numerical integration methods.
This way we will encounter, among other things, some novel (warped) $\textrm{AdS}_{3}$ solutions. Finally, we will discuss the eleven-dimensional origin of the various aforementioned solutions.
Some technical material concerning conventions for 7D spinors as well as the details of some more complicated flow equations will be collected in the appendices.

%%%%%%%%%%%%%%%%%%%%%%%%%%%%%%%%%%%%%%%%%%%%%%%%%%%%%%%%%
%%
%%               Main body
%%
%%%%%%%%%%%%%%%%%%%%%%%%%%%%%%%%%%%%%%%%%%%%%%%%%%%%%%%%%

\section{Minimal gauged supergravities in $D=7$}
\label{sec:gauged_sugra}

$\mathcal{N}=1$ (ungauged) supergravity in seven dimensions coupled to three vector multiplets can be obtained by reducing type I supergravity in ten dimensions on a $\mathbb{T}^3$. 
The theory possesses $16$ supercharges which can be rearranged into a pair of symplectic-Majorana (SM) spinors transforming as a doublet of $\textrm{SU}(2)_{R}$. 
In this paper we shall restrict to its minimal incarnation obtained as a truncation to the gravity supermultiplet. 

In this case, the full Lagrangian enjoys a global symmetry given by
\be
G_{0} \ = \ \mathbb{R}^{+}_{X} \, \times \, \textrm{SO}(3) \ .
\notag
\ee
The $(40_{B} \ + \ 40_{F})$ bosonic and fermionic propagating degrees of freedom (dof's) of the theory are then rearranged into irrep's of $G_{0}$ as described in table~\ref{Table:dofs}. We refer to 
the appendix for a summary of our notations concerning SM spinors.
\begin{table}[h!]
\renewcommand{\arraystretch}{1}
\begin{center}
\scalebox{1}[1]{
\begin{tabular}{|c|c|c|c|c|}
\hline
fields & $\textrm{SO}(5)$ irrep's & $\mathbb{R}^{+} \, \times \, \textrm{SO}(3)$ irrep's & $\textrm{SU}(2)_{R}$ irrep's & \# dof's  \\
\hline \hline
${e_{\mu}}^{m}$ & $\textbf{14}$ & $\textbf{1}_{(0)}$ & $\textbf{1}$ & $14$ \\
\hline
${A_{\mu}}^{i}$ & $\textbf{5}$ & $\textbf{3}_{(+1)}$ & $\textbf{1}$ & $15$ \\
\hline
$B_{\m\n\r}$ & $\textbf{10}$ & $\textbf{1}_{(-2)}$ & $\textbf{1}$ & $10$ \\
\hline
$X$ & $\textbf{1}$ & $\textbf{1}_{(+1)}$ & $\textbf{1}$ & $1$ \\
\hline
\hline
$\psi_{\m\a}$ & $\textbf{16}$ & $\textbf{1}_{(0)}$ & $\textbf{2}$ & $32$ \\
\hline
${\chi}_{\a}$ & $\textbf{4}$ & $\textbf{1}_{(0)}$ & $\textbf{2}$ & $8$ \\
\hline
\end{tabular}
}
\end{center}
\caption{{\it The on-shell field content of (ungauged) minimal $\mathcal{N}=1$ supergravity in $D=7$. Each field is massless and hence transforms in some irrep of the corresponding little group $\textrm{SO}(5)$ w.r.t.
spacetime diffeomorphisms and local Lorentz transformations.}} \label{Table:dofs}
\end{table}
In such a minimal setup, the possible consistent deformations of the theory associated with a generalized embedding tensor are of the following two different types:
\begin{itemize}
\item an $\textrm{SU}(2)$ gauging realized by the three vector fields in table~\ref{Table:dofs} and controlled by the gauge coupling constant $g$,
\item a St\"uckelberg-like coupling $h$ giving a mass to the 3-form gauge potential $B_{(3)}$ in the gravity multiplet. 
\end{itemize}
The bosonic Lagrangian for the deformed theory then reads \cite{Townsend:1983kk}
\be
\label{Lagrangian_7D}
\begin{array}{lclc}
\mathcal{L} & = & \mathcal{R}\,*_{(7)}1\,-\,5\,X^{-2}\,*_{(7)}dX\,\wedge\,dX \,-\,\dfrac{1}{2}\,X^{4}\,*_{(7)}\mathcal{F}_{(4)}\,\wedge\,\mathcal{F}_{(4)} \,-\, \mathcal{V}(X)\,*_{(7)}1   & \\[1mm]
  & & - \,\dfrac{1}{2}\,X^{-2}\,*_{(7)}\mathcal{F}_{(2)}^{i}\,\wedge\,\mathcal{F}_{(2)}^{i} \,-\, h\,\mathcal{F}_{(4)}\,\wedge\,B_{(3)} \,+\, \dfrac{1}{2}\,\mathcal{F}_{(2)}^{i}\,\wedge\,\mathcal{F}_{(2)}^{i} \,\wedge\,B_{(3)} & ,
\end{array}
\ee
where $\mathcal{R}$ denotes the 7-dimensional Ricci scalar, $\mathcal{V}(X)$ is the scalar potential and $\mathcal{F}_{(2)}$ \& $\mathcal{F}_{(4)}$ are the (modified) field strengths of the 1- and 3-form gauge potentials, respectively. 
Their explicit form is given by
\be
\begin{array}{lclc}
\mathcal{F}_{(2)}^{i} \ = \ dA^{i} \, - \, \dfrac{g}{2} \, \epsilon^{ijk}\,A^{j}\wedge A^{k} & \textrm{ and } & \mathcal{F}_{(4)} \ = \ dB_{(3)} & .
\end{array}
\ee
The explicit form of the scalar potential induced by the two aforementioned deformations reads
\be
\mathcal{V}(X) \ = \ 2h^{2} \, X^{-8} \, - \, 4\sqrt{2}\,gh\, X^{-3} \, - \, 2g^{2} \,X^{2} \ ,
\label{potential}
\ee
which may be, in turn, rewritten in terms of a real \emph{superpotential} 
\be
\label{superpotential}
f(X) \ = \ \frac{1}{2}\,\left(h\,X^{-4}\,+\,\sqrt{2}\,g\,X\right) \ ,
\ee
through the relation
\be
\mathcal{V}(X) \ = \ \frac{4}{5}\,\left(-6f(X)^{2}\,+\,X^{2}\,\left(D_{X}f\right)^{2}\right) \ .
\ee

Finally, due to the presence of the topological term in \eqref{Lagrangian_7D} induced by $h$ and $B_{(3)}$, one has to impose an odd-dimensional self-duality condition \cite{Townsend:1983xs} of the form
\be
\label{SD_cond}
X^{4} \, *_{(7)}\mathcal{F}_{(4)}\ \overset{!}{=} \ -2h \, B_{(3)} \, + \, \frac{1}{2} \, A^{i} \, \wedge \, \mathcal{F}_{(2)}^{i} \,+ \, 
\frac{g}{12} \epsilon_{ijk} \, A^{i} \, \wedge \, A^{j} \, \wedge \, A^{k} \ .
\ee

This supergravity theory enjoys $\mathcal{N}=1$ supersymmetry, which can be made manifest by checking the invariance of its full Lagrancian w.r.t. the following supersymmetry transformations
\be
\label{SUSY_eqns_7D}
\begin{array}{lclc}
\delta_{\zeta}{e_{\m}}^{m} & = & \bar{\zeta}_{a}\,\g^{m}\,{\psi_{\m}}^{a} & , \\[2mm]
\delta_{\zeta}X & = & \frac{X}{2\sqrt{10}}\,\bar{\zeta}_{a}\,\chi^{a} & , \\[2mm]
\delta_{\zeta}{A_{\m}}^{i} & = & i\,\frac{X}{\sqrt{2}} \,\left[\left(\bar{\psi}_{\m b}\,\zeta^{a}\,-\,\frac{1}{2}\,\d^{a}_{b}\,\bar{\psi}_{\m c}\,\zeta^{c}\right)
\,-\,\frac{1}{\sqrt{5}}\,\left(\bar{\chi}_{b}\,\g_{\m}\,\zeta^{a}\,-\,\frac{1}{2}\,\d^{a}_{b}\,\bar{\chi}_{c}\,\g_{\m}\,\zeta^{c}\right)\right] & , \\[2mm]
\delta_{\zeta}{B_{\m\n\r}} & = & \frac{X^{-2}}{\sqrt{2}} \, \left(\frac{3}{2}\,\bar{\psi}_{\m a}\,\g_{\n\r}\,\zeta^{a}
\,+\,\frac{1}{\sqrt{5}}\,\bar{\chi}_{a}\,\g_{\m\n\r}\,\zeta^{a}\right) & , \\[2mm]
\delta_{\zeta}{\psi_{\m}}^{a} & = & \nabla_{\m}\zeta^{a} \, + \, ig\,{\left(A_{\m}\right)^{a}}_{b}\,\zeta^{b}  \, + \, 
i\,\frac{X^{-1}}{10\sqrt{2}}\,\left({\g_{\m}}^{mn}\,-\,8\,{e_{\m}}^{m}\,\g^{n}\right) \, {\left(\mathcal{F}_{(2)\, mn}\right)^{a}}_{b}\,\zeta^{b}  &  \\[1mm]
&  & + \,\frac{X^{2}}{160}\,\left({\g_{\m}}^{mnpq}\,-\,\frac{8}{3}\,{e_{\m}}^{m}\,\g^{npq}\right) \, \mathcal{F}_{(4)\, mnpq}\,\zeta^{a} \, - \, \frac{1}{5}\,f(X)\,\g_{\m}\,\zeta^{a} & , \\[2mm]
\delta_{\zeta}{\chi^{a}} & = &  \frac{\sqrt{5}}{2}\,X^{-1}\slashed{\partial}X \, \zeta^{a} \, - \, i\, \frac{X^{-1}}{\sqrt{10}}\,  {\left(\slashed{F}_{(2)}\right)^{a}}_{b}\,\zeta^{b} 
\, + \, \frac{X^{2}}{2\sqrt{5}}\,  \slashed{F}_{(4)}\,\zeta^{a} \, - \, \frac{X}{5} \, D_{X}f \, \zeta^{a} & ,
\end{array}
\ee
where we introduced the following notation $\slashed{\omega}_{(p)}\,\equiv\,\frac{1}{p!}\,\g^{m_{1}\cdots m_{p}}\,\omega_{(p)\,m_{1}\cdots m_{p}}$, $\omega_{(p)}$ being a $p$-form, and the 
$\textrm{SU}(2)$-valued vector fields read
\be
A \ \equiv \ \frac{1}{2} \, A^{i} \, \s^{i} \ ,
\ee
$\left\{\sigma^{i}\right\}$ being the Pauli matrices given in \eqref{Pauli}.

The bosonic field equations obtained by varying the action \eqref{Lagrangian_7D} are given by  
\be
\label{EOM_7D}
\begin{array}{lclc}
\mathcal{R}_{\m\n} \, - \, 5X^{-2}\,\partial_{\m}X\,\partial_{\n}X \, - \, \frac{1}{5} \, \mathcal{V}(X) \, g_{\m\n} \, - \, \frac{X^{-2}}{2} \, \left(\mathcal{F}_{(2)}\right)^{2}_{\m\n}
 \, - \, \frac{X^{4}}{2} \, \left(\mathcal{F}_{(4)}\right)^{2}_{\m\n} & = & 0 & , \\[2mm]
\nabla_{\m}\left(X^{-1}\,\nabla^{\m}X\right) \, - \, \frac{X^{4}}{120}\,\left|\mathcal{F}_{(4)}\right|^{2} \, + \, \frac{X^{-2}}{20}\,\left|\mathcal{F}_{(2)}\right|^{2} \, - \, \frac{X}{10}\,D_{X}\mathcal{V} & = & 0 & , \\[2mm]
d\left(X^{4}\,*_{(7)}\mathcal{F}_{(4)}\right) \, + \, \frac{g}{\sqrt{2}}\,\mathcal{F}_{(4)} \, - \, \frac{1}{2} \, \mathcal{F}_{(2)}^{i} \, \wedge \, \mathcal{F}_{(2)}^{i} & = & 0 & , \\[2mm]
d_{A}\left(X^{-2}\,*_{(7)}\mathcal{F}_{(2)}^{i}\right) \, - \, \mathcal{F}_{(2)}^{i} \, \wedge \, \mathcal{F}_{(4)} & = & 0 & , \\[2mm]
\end{array}
\ee
where 
\be
\begin{array}{lcclc}
\left(\mathcal{F}_{(2)}\right)^{2}_{\m\n} \ \equiv \ {\mathcal{F}_{(2)}^{i}}_{\m\r}\,{{\mathcal{F}_{(2)}^{i}}_{\n}}^{\r} \, - \, \frac{1}{10}\,\left|\mathcal{F}_{(2)}\right|^{2} \, g_{\m\n} & , & & 
\left|\mathcal{F}_{(2)}\right|^{2} \ \equiv \ {\mathcal{F}_{(2)}^{i}}_{\m\n}\,{{\mathcal{F}_{(2)}^{i}}}^{\m\n} & , \\[2mm]
\left(\mathcal{F}_{(4)}\right)^{2}_{\m\n}  \ \equiv \ {\mathcal{F}_{(4)}}_{\m\r\s\kappa}\,{{\mathcal{F}_{(4)}}_{\n}}^{\r\s\kappa} \, - \, \frac{1}{40}\,\left|\mathcal{F}_{(4)}\right|^{2} \, g_{\m\n} & , & & 
\left|\mathcal{F}_{(4)}\right|^{4} \ \equiv \ {\mathcal{F}_{(4)}}_{\m\n\r\s}\,{{\mathcal{F}_{(4)}}}^{\m\n\r\s} & ,
\end{array}
\ee
and $d_{A}$ denotes the gauge-covariant differential.
Note that the equations of motion in \eqref{EOM_7D} are implied by the $\mathrm{SUSY}$ conditions \eqref{SUSY_eqns_7D} written down for a purely bosonic background.

\section{Domain wall solutions}

We first start reviewing domain wall (DW) solutions as special examples of supersymmetric solutions of this theory.
Supersymmetric DW's are BPS flows where the only excited degrees of freedom are the metric and the scalar fields. 
In this case we consider the following \emph{Ansatz} for the $D=7$ fields
\be
\begin{array}{lcll}
ds_{7}^{2} & = & e^{2U(r)}\,ds_{\textrm{Mkw}_{6}}^{2} \, + \, e^{2V(r)} \, dr^{2} & , \\[2mm]
X & = & X(r) & ,
\label{DWansatz}
\end{array}
\ee
where $ds_{\textrm{Mkw}_{6}}^{2}$ denotes the flat $\textrm{Mkw}_{6}$ metric,
while both vector and 3-form fields are kept vanishing. Note that the arbitrary function $V(r)$ is in fact non-dynamical and could be set to zero by means of a suitable gauge choice. 
However, when solving this type of problems, it is often convenient to keep such a gauge freedom in order to simplify the resulting flow equations such in way that they may be integrated analytically.

By choosing a Killing spinor of the form
\be
\zeta(r) \, = \, Y(r)\,\zeta_{0} \ ,
\ee
where $\zeta_{0}$ is a constant SM spinor (\emph{i.e.} obeying \eqref{SM_cond}) and further satisfying the following projection condition\footnote{In what follows we shall adopt the following notation: 
$\Pi(\mathcal{O}) \, \equiv \, \frac{1}{2} \, (\mathds{1}+\mathcal{O})$, where $\mathcal{O}$ denotes an idempotent spinorial operator.}
\be
\Pi(\g^{6}) \, \zeta_{0} \, = \, \zeta_{0} \ , 
\ee
the $\mathrm{SUSY}$ equations are fully implied by the following first-order flow equations
\be
\begin{array}{lclclc}
U' \, = \, \frac{2}{5} \, e^{V} \, f & , & Y' \, = \, \frac{Y}{5} \, e^{V} \, f & , & X' \, = \, -\frac{2}{5} \, e^{V} \,X^{2} \,D_{X}f & .
\label{eqDW}
\end{array}
\ee
If we make the gauge choice $e^{V}=-\frac{5\, X^{-2}}{2\, D_{X}f}\,$, the general solution of \eqref{eqDW} is given by
\be 
e^{2U}= \left(\frac{r}{\sqrt{2}\, g \, r^5-4 \,h}\right)^{1/2}\ , \qquad X=r\ ,
\label{DWflow}
\ee
where one can further consider special cases where $g\,=\,0$ \& $h\,\neq\,0$, $g\,\neq\,0$ \& $h\,=\,0$, and $g\,\neq\,0$ \& $h\,\neq\,0$. We will discuss their different 11D origin later in 
section~\ref{sec:Mtheory_lifts}.

\section{BPS flows with the 3-form potential}
\label{sec:flows}

In this section we present new classes of BPS flows within minimal gauged supergravity in $D=7$.
In particular we consider a new class of flows involving a non-trivial profile for the 3-form potential $B_{(3)}$, while still keeping the vectors inactivated for the moment. 
Note that in this case, one has two crucially different possibilities:
\begin{itemize}
\item  \textbf{Vanishing topological mass:} for these models the self-duality condition \eqref{SD_cond} is trivially satisfied by an electric profile for the 3-form potential and this cases are well described by the
already known membrane solutions of ungauged supergravity.

\item  \textbf{Non-vanishing topological mass:} for these models the condition in \eqref{SD_cond} requires a more complicated ``dyonic'' \emph{Ansatz} for the 3-form potential. This is the new situation 
that we analyze here and this will give rise to BPS solutions with 8 real supercharges ($\mathrm{BPS}$/2) \footnote{This situation was originally considered in \cite{Liu:1999ai}, where some insights were
given concerning the search for dyonic membrane solutions. However, at least to our knowledge, explicit solutions of this type have not been constructed yet.}. 
\end{itemize}

\subsection{Charged flow on the background $\mathrm{Mkw}_{3}\times \mathbb{R}^3$}

Let us now consider a non-trivial dyonic profile for the 3-form potential $B_{(3)}$ for a 7-dimensional background including the flat manifold $\mathbb{R}^{3}$.
We make the following \emph{Ansatz} for the fields
\be
\begin{array}{lcll}
ds_{7}^{2} & = & e^{2U(r)}\,ds_{\mathrm{Mkw}_{3}}^{2} \, + \, e^{2V(r)} \, dr^{2} \, + \, e^{2W(r)}\,ds_{\mathbb{R}^3}^{2} & , \\[2mm]
X & = & X(r) & , \\[2mm]
B_{(3)} & = & k(r)\,\textrm{vol}_{\mathrm{Mkw}_{3}} \, + \, l(r)\,\textrm{vol}_{\mathbb{R}^3} & ,
\end{array}
\ee
where $ds_{\mathrm{Mkw}_{3}}^{2}$ \& $ds_{\mathbb{R}^{3}}^{2}$ respectively denote the flat $\textrm{Mkw}_{3}$ \& the flat $\mathbb{R}^{3}$ metric,
while the vector fields are still kept vanishing. Note that $V(r)$ is an arbitrary non-dynamical function 
and can be set to zero with a suitable gauge choice. 

By choosing a Killing spinor of the form
\be
\zeta(r) \, = \, Y(r)\,\left(\cos\theta(r)\,\mathds{1}_{8}\,+\,\sin\theta(r)\,\g^{012}\right)\,\zeta_{0} \ ,
\label{Kspinor}
\ee
where $\zeta_{0}$ is a constant SM spinor (\emph{i.e.} obeying \eqref{SM_cond}) and further satisfying the following projection condition
\be
\Pi(\g^{3}) \, \zeta_{0} \, = \, \zeta_{0} \ , 
\label{gamma3proj}
\ee
the Killing spinor equations are fully implied by the following first-order flow equations
\be\left\{
\begin{array}{lclclc}
U' & = & \frac{1}{5} \, e^{V} \, f \, \frac{(3 \cos(4 \theta)\,-\,1)}{\cos(2 \theta)}  & , \\[2mm]
W' & = & -\frac{2}{5} \, e^{V} \, f \, \frac{(\cos (4 \theta)\,-\,2)}{\cos(2 \theta)}  & , \\[2mm]
Y' & = & \frac{1}{10} \, e^{V} \, Y \, f \, \frac{(3 \cos (4 \theta)\,-\,1)}{\cos(2 \theta)}  & , \\[2mm]
\theta' & = & -e^{V} \, f \, \sin (2 \theta) & , \\[2mm] 
k' & = & -\frac{4 f\, e^{3 U+V}}{X^2} \, \tan (2 \theta) & , \\[2mm] 
l' & = & \frac{4 f\, e^{V+3 W}}{X^2} \, \sin (2 \theta) & , \\[2mm] 
X' & = & -\frac{2}{5} \, e^{V} \, X \, \left(X\, D_{X}f\,-\,8 f\, \frac{\sin^4\theta}{\cos (2 \theta)}\right) & ,
\end{array}\right.
\label{floweqDW}
\ee
provided that the following extra differential constraint
\be
\label{constraintDW}
X \, D_{X}f \, + \, 4\,f \, \overset{!}{=} \, 0 \ ,
\ee
holds along the flow.
It can be shown that \eqref{constraintDW} is solved by a superpotential 
of the original form given in \eqref{superpotential} by setting $g\,=\,0$. This situation corresponds to having a pure St\"uckelberg deformation associated with the 
parameter $h$, without any gauging.

After performing the following gauge choice for the function $V$
\be
e^{V} \, \overset{\textrm{gauge fix.}}{=} \, f^{-1} \ ,
\ee
the above flow equations may be integrated analytically and the solution reads
\be
\label{ch_DW_noAdS}
\begin{array}{lclclc}
e^{2U} \, = \, \sinh(4r)^{1/5}\,\coth(2r) & & , & & e^{2V} \, = \frac{4}{h^2}\, \sinh(4r)^{16/5} & , \\[2mm]
e^{2W} \, = \, \sinh(4r)^{1/5}\,\tanh(2r) & & , & & X \, = \, \sinh(4r)^{2/5} & , \\[2mm]
k \, = \, \frac{1}{\sqrt{2}\,\sinh^{2}(2r)} & & , & & l \, = \, -\frac{1}{\sqrt{2}\,\cosh^{2}(2r)} & , \\[2mm]
Y \, = \, \sinh(4r)^{1/20}\,\coth(2r)^{1/4} & & , & & \theta \, = \, \arctan \left(e^{-2r}\right) & .
\end{array}
\ee
One may check that \eqref{ch_DW_noAdS} correctly satisfies the bosonic field equations in \eqref{EOM_7D} as well as the odd-dimensional self-duality condition \eqref{SD_cond}. Note that this solution is
not asymptotically $\textrm{AdS}_{7}$, consistently with the fact that the monomial scalar potential induced by the only contribution of the topological mass has a run-away behavior in $X$. 

\subsection{Charged flow on the background $\mathrm{Mkw}_{3}\times S^3$}

It is now natural to wonder if flows driven by the complete profile of the potential \eqref{potential} exist or, equivalently, if asymptotically $\mathrm{AdS}_7$ solutions with a running profile for the 3-form exist in the considered theory.

It is well known that 
one of the main features of the first-order formulation of supergravity is its gauge-dependence: the profile
of the Killing spinor directly determines the background through the first-order flow equations, which turn out to explicit depend on the spin connection of the background itself. Adapting this story to 
our case, this implies that searching for 
 Killing spinors corresponding to asymptotically $\mathrm{AdS}_7$ flows is equivalent to looking for a background parametrization such that the corresponding flow equations are
driven by the complete superpotential \eqref{superpotential}. 

We claim that this happens only if the locally Euclidean part of the background admits an $\mathrm{SO(3)}$-covariant {\itshape parallelized basis}, \emph{i.e.} we need a
field configuration parametrized in such a way the spin connection of the Euclidean part of the metric takes non-zero constant values once expressed in flat coordinates.
From these considerations it follows that the presence of $\mathrm{AdS}_7$ is excluded for a metric containing $\mathbb{R}^3$ since it is flat and also for $\mathbb{H}^3$ since its parallelized basis is $\mathrm{SO(2,1)}$-covariant.

Thus we consider an \emph{Ansatz} of the form,
\be
\begin{array}{lcll}
ds_{7}^{2} & = & e^{2U(r)}\,ds_{\textrm{Mkw}_{3}}^{2} \, + \, e^{2V(r)} \, dr^{2} \, + \, e^{2W(r)}\,ds_{S^{3}}^{2} & , \\[2mm]
X & = & X(r) & , \\[2mm]
B_{(3)} & = & k(r)\,\textrm{vol}_{\textrm{Mkw}_{3}} \, + \, l(r)\,\textrm{vol}_{S^{3}} & ,
\label{ansatz:mkw21}
\end{array}
\ee
where $ds_{S^{3}}^{2}$ is the metric of a unit $S^{3}$ and $\text{vol}_{S^{3}}$ its volume.
We choose the set of Hopf coordinates $(\theta_1,\theta_2, \theta_3)$ on $S^3$, such that
\be
ds_3^2=\frac{1}{\kappa^2}\left[ d\theta_2^2+\cos^2\theta_2 d\theta_3^2+\left(d\theta_1+\sin\theta_2d\theta_3\right)^2   \right] \, .
\label{S3}
\ee
The \emph{dreibein} corresponding to this parametrization of the $S^3$ is non-diagonal,
\be
 \begin{array}{lclc}
e^{1} & = & \dfrac{1}{\kappa} \,d\theta_{1} & ,\\[2mm]
e^{2} & = & \dfrac{1}{\kappa} \, \left(\cos\theta_{1}d\theta_{2}\,+\,\sin\theta_{1}d\theta_{3}\right) & ,\\[2mm]
e^{3} & = & \dfrac{\sin\theta_{2}}{\kappa} \,d\theta_{1}\,+\,\dfrac{\cos\theta_{2}}{\kappa} \,\left(-\sin\theta_{1}d\theta_{2}\,+\,\cos\theta_{1}d\theta_{3}\right)& ,
\end{array}
\label{S3vielbein}
\ee
and the corresponding spin connection is constant if expressed in the flat basis \eqref{S3vielbein} and given by
\be
\omega_{i\,jk}=\frac{\kappa}{2}\,\epsilon_{ijk} \quad \text{ with } \quad i,j,k=1,2,3\,.
\label{spinconnS3}
\ee
In what follows the $\mathrm{\mathrm{SO(3)}}$ indices $(i,j,k)=1,2,3$ must be identified with the $(4,5,6)$ components of the flat basis of the whole 7-dimensional metric.

By choosing a Killing spinor with the same profile of \eqref{Kspinor} and satisfying 
the projection condition \eqref{gamma3proj}, the Killing spinor equations are satisfied if the following 
system of first-order flow equations hold,
\be\left\{
\begin{array}{lclclc}
U' & = & \frac{1}{25} \, e^{V} \, \frac{(3 \cos(4 \theta)\,+\,7)\, f+6\sin^2(2 \theta)\, X\, D_{X}f}{\cos(2 \theta)}  & , \\[2mm]
W' & = &  -\frac{2}{25} \, e^{V} \, \frac{(\cos(4 \theta)\,-\,6)\, f+2\sin^2(2 \theta)\, X\, D_{X}f}{\cos(2 \theta)}  & , \\[2mm]
Y' & = & \frac{1}{50} \, e^{V} \,Y\, \frac{(3 \cos(4 \theta)\,+\,7)\, f+6\sin^2(2 \theta)\, X\, D_{X}f}{\cos(2 \theta)}  & , \\[2mm]
\theta' & = & -\frac15 \,e^{V} \,  \sin (2 \theta)\,\left(f-\, X\, D_{X}f\right) & , \\[2mm] 
k' & = & \frac25 \frac{e^{V+3 U}\tan(2\theta)\,\left(2\,f+3\, X\, D_{X}f\right)}{X^2} & , \\[2mm] 
l' & = & \frac45 \frac{e^{V+3 W}\sin(2\theta)\,\left(f-\, X\, D_{X}f\right)}{X^2} & , \\[2mm] 
X' & = & -\frac{2}{25} \, e^{V} \, X \,\frac{(4+\cos(4\theta))\,X\, D_{X}f\,+\,2 \sin^2(2 \theta)\,f}{\cos (2 \theta)} & ,
\end{array}\right.
\label{floweqAdSDW}
\ee
where the constraint
\be
\label{constraintAdSDW}
\kappa+\frac25 e^{W}\tan(2\theta)\left(X \, D_{X}f \, + \, 4\,f \right)\overset{!}{=} \, 0 \ ,
\ee
has been imposed along the whole flow. Imposing the constraint \eqref{constraintAdSDW} and the flow
\eqref{floweqAdSDW} on the equations of motion \eqref{EOM_7D}, it follows that they are satisfied if the superpotential is given by the 
\eqref{superpotential} with arbitrary values of $g$ and $h$.

As in the case of the previous section, we perform the following gauge choice for the function $V$
\be
e^{-V} \, \overset{\textrm{gauge fix.}}{=} \, \frac15 \left (f\,-\,X\,D_Xf\right) \ ,
\label{gauge}
\ee
the above flow equations may be integrated analytically and the solution reads
\be
\label{ch_DW_AdS}
\begin{array}{lclclc}
e^{2U} \, = \,\left(\frac{\left(\rho ^4+1\right)^3 \left(\rho ^{16}+4 \rho ^{12}+4 \rho ^4+1\right)}{\rho ^{10}
   \left(\rho ^4-1\right)^2}\right)^{2/5} & & , & & e^{2V}\,=\, \frac{ 2^{2/5} \left(\rho ^8-1\right)^{16/5}}{h^{2/5} g^{8/5} \left(\rho
   ^{16}+4 \rho ^{12}+4 \rho ^4+1\right)^{8/5}} & , \\[2mm]
e^{2W} \, = \, \left(\frac{\left(\rho ^4-1\right)^3 \left(\rho ^{16}+4 \rho ^{12}+4 \rho ^4+1\right)}{\rho ^{10}
   \left(\rho ^4+1\right)^2}\right)^{2/5} & & , & & X \, = \, \frac{2^{3/10} h^{1/5} \left(\rho ^8-1\right)^{2/5}}{
   g^{1/5} \left(\rho ^{16}+4 \rho
   ^{12}+4 \rho ^4+1\right)^{1/5}} & , \\[2mm]
k \, = \, \frac{2^{2/5} g^{2/5} \left(\rho ^{16}+4 \rho ^{12}+4 \rho ^4+1\right)}{h^{2/5} \rho ^4
   \left(\rho ^4-1\right)^2} & & , & & l \, = \,\frac{2^{2/5} g^{2/5} \left(\rho ^{16}-4 \rho ^{12}-4 \rho ^8-4 \rho ^4+1\right)}{h^{2/5}
   \rho ^4 \left(\rho ^4+1\right)^2} & , \\[2mm]
Y \, = \, \left(\frac{\left(\rho ^4+1\right)^3 \left(\rho ^{16}+4 \rho ^{12}+4 \rho ^4+1\right)}{\rho ^{10}
   \left(\rho ^4-1\right)^2}\right)^{1/10} & & , & & \theta \, = \, \arctan \left(\rho^{-2}\right) & ,
\end{array}
\ee
where $r=\log\,\rho$ and from \eqref{constraintAdSDW} one obtains $\kappa=- 2^{9/5} g^{4/5} h^{1/5}$.

In the asymptotic region, the flow \eqref{ch_DW_AdS} turns out to locally reproduce $\mathrm{AdS_7}$, in fact the 
contribution of $\mathcal{F}_{(4)}$ turns out to be sub-leading when $r \rightarrow +\infty$. 
In this limit one has
\be
\theta=0\, ,\quad X= 1\,, \quad \mathcal{F}_{0123}=0\,, \quad \mathcal{F}_{3456}=0\, ,
\ee
where we made the choice for the parameters\footnote{The explicit dependence on the parameters $h\text{ and }g$
of the flow is related to the gauge choice \eqref{gauge}.
Given this particular gauge choice, one can always choose $h=\frac{g}{2\sqrt{2}}$ in order to obtain $X=1$ as an asymptotic of
value for the $\mathbb{R}^{+}$ dilaton.} $h$ and $g$ such that $h=\frac{g}{2\sqrt{2}}$.
In the limit $r \rightarrow 0$ the flow is singular. Finally it is easy to verify that \eqref{ch_DW_AdS} correctly satisfies the equations of motion
in \eqref{EOM_7D} and the odd-dimensional self-duality condition \eqref{SD_cond}.

\subsection{Background $\mathrm{AdS}_3\times S^3$: $\mathrm{AdS}_7$ charged domain wall}

We now want to consider a slightly more complicated system such that the whole background be curved. This is achieved by considering an $\mathrm{AdS}_3$ slicing of the 7-dimensional background. In this section we will consider
for simplicity a background depending only on a independent warp factor $U(r)$, thus the configuration of the fields has the form,
\be
\begin{array}{lcll}
ds_{7}^{2} & = & e^{2U(r)}\,ds_{\mathrm{AdS}_3}^{2} \, + \, e^{2V(r)} \, dr^{2} \, + \, e^{2U(r)}\,ds_{S^{3}}^{2} & , \\[2mm]
X & = & X(r) & , \\[2mm]
B_{(3)} & = & k(r)\,\textrm{vol}_{\mathrm{AdS}_3} \, + \, l(r)\,\textrm{vol}_{S^{3}} & ,
\end{array}
\ee
where $ds_{S^{3}}^{2}$ is again the metric of the $S^3$ parametrized as in \eqref{S3}, while $ds_{\mathrm{AdS}_3}^{2}$ is the metric of $ds_{\mathrm{AdS}_3}^{2}$ in the parallelized basis $(t,x^1,x^2)$ such that
\be
ds_{\mathrm{AdS}_3}^{2}=\frac{1}{L^2}\left[ (dx^1)^2+\cosh^2x^1 (dx^2)^2-\left(dt-\sinh x^1 d x^2\right)^2   \right] \, .
\label{AdS3}
\ee
The non-symmetric \emph{dreibein} associated to this parametrization is given by
\be
 \begin{array}{lclc}
e^{0} & = & \dfrac{1}{L} \,dt & ,\\[2mm]
e^{1} & = & \dfrac{1}{L} \, \left(\cos t \,dx^{1}\,+\,\sin t \,dx^{2}\right) & ,\\[2mm]
e^{2} & = & -\dfrac{\sinh x^{1}}{L} \,dt\,+\,\dfrac{\cosh x^{1}}{L} \,\left(-\sin t \,dx^{1}\,+\,\cos t \,dx^{2}\right)& ,
\end{array}
\label{AdS3vielbein}
\ee
and defines a constant spin connection as in the case of $S^3$.

Keeping the same Killing spinor given in \eqref{Kspinor} with 
the projection condition \eqref{gamma3proj}, the Killing spinor equations determine a
system of first-order flow equations for the superpotential \eqref{superpotential} if 
\be
\theta(r)=0\ ,\quad k(r)=l(r)\ ,\quad \kappa=L\ .
\ee
In this case the BPS equations take the simple form
\be\left\{
\begin{array}{lclclc}
U' \, = \, \frac{2}{5} \, e^{V} \, f  & , \\[2mm]
Y'  \, = \, \frac{Y}{5} \, e^{V} \, f & , \\[2mm]
k' \, = - \frac{e^{2U+V}\,L}{ X^2} & , \\[2mm] 
X' \, = \, -\frac{2}{5} \, e^{V} \,X^{2} \,D_{X}f & ,
\end{array}\right.
\label{floweqDW_U=W}
\ee
Choosing the gauge
\be
e^{-V}\,=\,-\frac25\,X^2\, D_{X}f\ ,
\ee
 and choosing the parameters as $h=\frac{g}{2\sqrt{2}}$, the equations \eqref{floweqDW_U=W} are easly integrated in the interval $r\in (0,1)$, yielding
 \be\left\{
\begin{array}{lclclc}
e^{2U}\, =  \dfrac{2^{-1/4}}{\sqrt{g}}\,\left(\dfrac{r}{1\,-\,r^{5}}\right)^{1/2}\,&, \\[2mm]
e^{2V} \, =  \dfrac{25}{2g^{2}}\,\dfrac{r^{6}}{\left(1\,-\,r^{5}\right)^{2}}\,& , \\[2mm]
Y \, =   \, \dfrac{2^{-1/16}}{g^{1/8}}\,\left(\dfrac{r}{1\,-\,r^{5}}\right)^{1/8}\,&, \\[2mm]
k \,=  -\dfrac{2^{1/4}\,L}{g^{3/2}}\left(\dfrac{r^{5}}{1\,-\,r^{5}}\right)^{1/2} \,& , \\[2mm]
X\, = \, r\,&. 
\label{AdS3_DW_AdS_U=V}
\end{array}\right.
\ee
This solution turns out to be asymptotically locally $\mathrm{AdS}_7$. In particular, in the limit $r \rightarrow 1$ one has
\be
 X= 1 \ , \quad \mathcal{F}_{0123}=0\ , \quad \mathcal{F}_{3456}=0\ ,
\ee
while for $r \rightarrow 0$ the solution is singular.

\subsection{Background $\mathrm{AdS}_3\times S^3$: general flow $\mathrm{AdS}_7 \rightarrow \mathrm{AdS}_3\times \mathbb{T}^4$}

Let us now consider a slightly more complicated background where the warping is determined by two independent functions $U$ and $W$,
\be
\begin{array}{lcll}
\label{backgroundAdS3}
ds_{7}^{2} & = & e^{2U(r)}\,ds_{\mathrm{AdS}_3}^{2} \, + \, e^{2V(r)} \, dr^{2} \, + \, e^{2W(r)}\,ds_{S^{3}}^{2} & , \\[2mm]
X & = & X(r) & , \\[2mm]
B_{(3)} & = & k(r)\,\textrm{vol}_{\mathrm{AdS}_3} \, + \, l(r)\,\textrm{vol}_{S^{3}} & ,
\end{array}
\ee
where $ds_{S^{3}}^{2}$ is again the metric of the $S^3$ parametrized as in \eqref{S3}, while $ds_{\mathrm{AdS}_3}^{2}$ is the metric of $ds_{\mathrm{AdS}_3}^{2}$ parametrized as in \eqref{AdS3}.

Given the usual Killing spinor \eqref{Kspinor} with 
the projection condition \eqref{gamma3proj}, the first-order flow equations are given by
\be\left\{
\begin{array}{lclclc}
U' & = & \frac{1}{25} \, e^{V} \, \frac{(3 \cos(4 \theta)\,+\,7)\, f+6\sin^2(2 \theta)\, X\, D_{X}f-\,5L\,e^{-U} \sin(2\theta)}{\cos(2 \theta)}  & , \\[2mm]
W' & = &  -\frac{1}{25} \, e^{V} \, \frac{2(\cos(4 \theta)\,-\,6)\, f+4\sin^2(2 \theta)\, X\, D_{X}f+\,5L\,e^{-U}\sin(2\theta)}{\cos(2 \theta)}  & , \\[2mm]
Y' & = & \frac{1}{50} \, e^{V} \,Y\, \frac{(3 \cos(4 \theta)\,+\,7)\, f+6\sin^2(2 \theta)\, X\, D_{X}f-\,5L\,e^{-U} \sin(2\theta)}{\cos(2 \theta)}  & , \\[2mm]
\theta' & = & -\frac15 \,e^{V} \,  \sin (2 \theta)\,\left(f-\, X\, D_{X}f\right) & , \\[2mm] 
k' & = &  \frac{e^{3U+V}}{5\, X^2}\left[ 2 \tan(2\theta)\, \left(2f+\,3\, X\, D_{X}f\right) -\frac{5\,L\,e^{-U}}{\cos(2\theta)} \right] & , \\[2mm] 
l' & = & \frac{e^{3W+V}}{5\, X^2}\left[ 4 \sin(2\theta)\, \left(f\,-\, X\, D_{X}f\right) -5\,L\,e^{-U} \right] & , \\[2mm] 
X' & = & -\frac{1}{25} \, e^{V} \, X \,\frac{2\,(4+\cos(4\theta))\,X\, D_{X}f\,+\,4 \sin^2(2 \theta)\,f-\,5\,L\,e^{-U}\sin(2\theta)}{\cos (2 \theta)} & ,
\end{array}\right.
\label{flowWarpedAdS}
\ee
where the constraint
\be
\label{constraintWarpedAdS}
\kappa-L\,\frac{e^{W-U}}{\cos(2\theta)}+\frac25 e^{W}\tan(2\theta)\left(X \, D_{X}f \, + \, 4\,f \right)\overset{!}{=} \, 0 \ ,
\ee
has been imposed along the whole flow.
Imposing the constraint \eqref{constraintWarpedAdS} the equations of motion \eqref{EOM_7D} are fully satisfied imposing \eqref{flowWarpedAdS} if the superpotential is given by the 
\eqref{superpotential} with arbitrary values of $g$ and $h$.

Performing the usual gauge choice for the function $V$
\be
e^{-V} \, \overset{\textrm{gauge fix.}}{=} \, \frac15 \left (f\,-\,X\,D_Xf\right) \ ,
\label{gauge}
\ee
the above flow equations are solved by
\be
\label{AdS3_DW_AdS}
\begin{array}{lclclc}
e^{2U} & = & \frac{\left(\rho ^4+1\right)^2 \left(\sqrt{2} \,g\, \left(\rho ^{16}+4\,
   \rho ^{12}+4\, \rho ^4+1\right)-8 \,L \,\rho ^4 \left(\rho
   ^8+1\right)\right)^{2/5}}{ 2^{14/5} \,h^{2/5}\, \rho ^4  \left(\rho
   ^8-1\right)^{4/5}}  & , \\[2mm] 
e^{2V} & = & \frac{ 2^{26/5}\, \left(\rho ^8-1\right)^{16/5}}{h^{2/5}
   \left(\sqrt{2}\,g\, \left(\rho ^{16}+4 \rho ^{12}+4 \rho
   ^4+1\right)-8 \,L \,\rho ^4\, \left(\rho ^8+1\right)\right)^{8/5}}  & , \\[2mm]
   
e^{2W}  & = &\frac{\left(\rho ^4-1\right)^2 \left(\sqrt{2}\, g\, \left(\rho
   ^{16}+4\, \rho ^{12}+4\, \rho ^4+1\right)-8\, L\, \rho ^4 \left(\rho
   ^8+1\right)\right)^{2/5}}{ 2^{14/5}\, h^{2/5} \rho ^{4}
   \left(\rho ^8-1\right)^{4/5}}  & , \\[2mm]  
X  & = & \frac{2^{2/5} \,h^{1/5}\,\left(\rho ^8-1\right)^{2/5}}{\left(\sqrt{2}\,
   g\, \left(\rho ^{16}+4\, \rho ^{12}+4\, \rho ^4+1\right)-8\, L\, \rho ^4
   \left(\rho ^8+1\right)\right)^{1/5}} & , \\[2mm] 
   
k  & = &\frac{\sqrt{2} \,g\, \left(\rho ^{16}+4\, \rho ^{12}+4 \rho ^4+1\right)-2\, L\, \left(\rho
   ^{16}+4 \rho ^{12}-2 \,\rho ^8+4\, \rho ^4+1\right)}{16\, h\, \rho ^4 \left(\rho ^4-1\right)^2}  & , \\[2mm]
  l  & = &\frac{\sqrt{2}\,g\, \left(-\rho ^{16}+4\, \rho ^{12}+4\, \rho ^8+4\, \rho ^4-1\right)+2\, L\,
   \left(\rho ^{16}-4\, \rho ^{12}-2\, \rho ^8-4\, \rho ^4+1\right)}{16 \,h\, \rho ^4 \left(\rho
   ^4+1\right)^2} & , \\[2mm]
Y  & = &\left(\frac{\left(\rho ^4+1\right)^2 \left(\sqrt{2} \,g\, \left(\rho ^{16}+4\,
   \rho ^{12}+4\, \rho ^4+1\right)-8 \,L \,\rho ^4 \left(\rho
   ^8+1\right)\right)^{2/5}}{ 2^{14/5} \,h^{2/5}\, \rho ^4  \left(\rho
   ^8-1\right)^{4/5}}\right)^{1/4} & , \\[2mm]
\theta  & = & \arctan \left(\rho^{-2}\right) & ,
\end{array}
\ee
where $r=\log\,\rho$ and, from \eqref{constraintWarpedAdS}, one obtains 
\label{k&Lrelation}
\be
\kappa+L\,=\,\sqrt{2}\,g\ .
\ee
This flow is asymptotically locally $\mathrm{AdS_7}$: for any values of $\kappa$ and $L$ respecting \eqref{k&Lrelation} and 
for $h=\frac{g}{2\,\sqrt{2}}$, one has 
\be
\theta=0\ ,\quad X= 1\ , \quad \mathcal{F}_{0123}=0\ , \quad \mathcal{F}_{3456}=0\  ,
\ee
in the limit $r \rightarrow +\infty$.

The study of the limit $r\rightarrow 0$ crucially depends on the relation between $\kappa$ and $L$. The general leading-order behavior of the scalar potential \eqref{potential} is given by
\be
\mathcal{V} \,=\, \frac{ h^{2/5}\, \left(5 \,\sqrt{2}\, g-8\, L\right)^{8/5}  }{ 2^{3/10}\, r^{16/5}}\,+\,\dots\ .
\ee
From this expression we conclude that the behavior of the flow in the limit $r \rightarrow 0$ is singular except for the special value
\be
L\,=\,\frac{5\,g}{4\,\sqrt{2}}\ ,
\label{magicrelation}
\ee
where the scalar potential takes a constant value and the flow turns out to be described locally by $\mathrm{AdS}_3\times \mathbb{T}^4$, where the main difference with respect to the asymptotics is the fact that this geometry is
not a solution per se, as $\mathrm{AdS}_7$, but only the infrared (leading) profile of the flow \eqref{AdS3_DW_AdS} when the radii of $\mathrm{AdS}_3$ and $S^3$ are related by \eqref{magicrelation}.

In this limit, we have
\be
\theta=\frac{\pi}{4}\  ,\quad X= \frac{2^{2/5} }{3^{1/5}}\ , \quad \mathcal{F}_{0123}=0\ , \quad \mathcal{F}_{3456}=-\frac{3^{1/5}}{2^{19/10} }\,g\  .
\ee
Finally one can verify that \eqref{AdS3_DW_AdS} solves the equations of motion \eqref{EOM_7D} and the odd-dimensional self-duality condition \eqref{SD_cond}.

\section{Coupling to the $\mathrm{\mathrm{SU(2)}}$ vectors}
\label{sec:vectors}

In this section we extend our analysis including the coupling to the $\mathrm{\mathrm{SU(2)}}$ vectors $A^i$. In particular, the aim is finding solutions described by the
backgrounds \eqref{ansatz:mkw21} and \eqref{backgroundAdS3}, with running 3-form field,
including three non-Abelian vectors describing a Hopf fibration of the 3-sphere $S^3$. Extending the set of excited fields in the general \emph{Ansatz} results in a partial supersymmetry breaking. 
On the one hand this is due to the 
presence of new terms in the Killing spinor equations \eqref{SUSY_eqns_7D}, on the other hand, the stucture of \eqref{SUSY_eqns_7D} tells how the profile of the vectors should be in order to still preserve some amount of supersymmetry. 

\subsection{Killing spinors and twisting condition}

Let us consider the backgrounds \eqref{ansatz:mkw21} or \eqref{backgroundAdS3} with the $S^3$ metric parametrized as in \eqref{S3}, together with an \emph{Ansatz} for the vectors given by
\be
A_j^i=\frac{A(r)}{2\,g}\,\epsilon^{i\,k\,l}\,\omega_{j\,kl}\ ,
\label{ansatzvector}
\ee
where $\omega_i$ are the components of the spin connection of the $S^3$ and the last three values of the curved index $\mu=4,5,6$ have been identified with the $\mathrm{SO(3)}$ indices $i,j\cdots=1,2,3$.

Given the \emph{Ansatz} \eqref{ansatzvector}, we notice that the SM structure of the spinors turns out to be crucial in order to avoid a complete SUSY breaking. This may be seen explicitly by looking at the
 gravitini supersymmetry variations $\delta_{\zeta}{\psi_{\m}}^{a}$, which acquire now the following new terms depending on $A^i$
\be
\cdots +\frac14\,\omega_{i\,\,jk}\,\gamma^{j\,k}\zeta^a + \, ig\,{\left(A_{i}\right)^{a}}_{b}\,\zeta^{b}+ \, 
i\,\frac{X^{-1}}{10\sqrt{2}}\,\left({\g_{i}}^{mn}\,-\,8\,{e_{i}}^{m}\,\g^{n}\right) \, {\left(\mathcal{F}_{mn}\right)^{a}}_{b}\,\zeta^{b}+\cdots\  ,
\label{susyvariationsvector}
\ee
which are characterized by a non-trivial action of the vectors on the $\mathrm{\mathrm{SU(2)}}$ structure of the spinor $\zeta^a$. If one looks at first contribution in \eqref{susyvariationsvector} 
coming from the spin connection of the $S^3$ in relation to the 
second term, we see that the only way of preserving some supersymmetry is to take the Killing spinor \emph{oriented along} the direction identified by the vectors. This happens only if one imposes three new projection condition on the spinor. 
In terms of the SM spinor $\zeta^a$ defined in \eqref{Kspinor} and satisfying \eqref{gamma3proj}, these new conditions are given by
\be
\gamma^{5\,6}\, \zeta_0^a=-i\,\left(\sigma^1 \right )^a_{\,\,\,b}\zeta_0^b\ ,\quad \gamma^{4\,6}\, \zeta_0^a=i\,\left(\sigma^2 \right )^a_{\,\,\,b}\zeta_0^b\ , \quad
 \gamma^{4\,5}\, \zeta_0^a=-i\,\left(\sigma^3 \right )^a_{\,\,\,b}\zeta_0^b\ ,
 \label{vectorproj}
\ee
which may be reexpressed as
\be
\Pi\left(i\,\g^{ij}\,\otimes\,\s^{k}\right)\,\zeta_0 \overset{!}{=} \, \zeta_0 \ ,
\ee
with $i,\,j,\,k$ chosen to be all different and in all possible permutations.
It easy to show that the SM condition \eqref{SM_cond} is given exactly by the second projection condition in \eqref{vectorproj} if one represents the spinor $\zeta_0^a$ as a $\mathrm{\mathrm{SU(2)}}$ doublet. 
Thus \eqref{vectorproj} reduce the total amount of supersymmetry to two real supercharges ($\mathrm{BPS}/8$).

It has been shown \cite{Acharya:2000mu} that the projection conditions \eqref{vectorproj} are naturally realized from those configurations with $A(r)=1$ and the gauge fields are independent of the radial coordinate.
In this case, the effect of the vectors \eqref{ansatzvector} is to exactly compensate
the contribution in \eqref{susyvariationsvector} due to the spin connection of $S^3$. This can be understood by recalling the expression of the spin connection of
$S^3$ given in \eqref{spinconnS3} and comparing the first two terms of \eqref{susyvariationsvector}. It is easy to show that \eqref{vectorproj} 
are implied by a \emph{twisting condition} \cite{Maldacena:2000mw} given by
\be 
-\frac12 \,\omega_{i\,\,jk}\,\gamma^{j\,k}\zeta^a=i\,g\,  A^j_i \,\left(\sigma^j \right )^a_{\,\,\,b}\zeta^b\ .
\label{twisting}
\ee
Thus, in this case, the effect of the coupling to the vector fields is literarly to twist the Killing spinor in order to compensate the contribution coming from the curvature of the background and preserving
a certain amount of supersymmetry. 

It is worth mentioning that including of the 3-form implies a non-trivial radial dependence for the gauge fields. In the next sections we will provide some examples of this fact. Generally the special form 
of the Killing spinor \eqref{Kspinor}, which is needed in order to include the 3-form, implies a non-trivial profile for $A(r)$ and from this it follows that all the solutions of the type $\mathrm{AdS}_{p}\times \Sigma_{7-p}$ are either
characterized by a constant value for $A(r)$ and a vanishing 4-form field strength, or by a non-constant profile for the gauge fields and a non-trivial the 3-form.

\subsection{Vectors coupled to the background $\mathrm{Mkw}_{3}\times S^3$}
\label{mkw21vector}

Let us consider the background \eqref{ansatz:mkw21}, and furthermore include vectors given by the \emph{Ansatz} \eqref{ansatzvector}. Thus one has
\be
\begin{array}{lcll}
ds_{7}^{2} & = & e^{2U(r)}\,ds_{\textrm{Mkw}_{3}}^{2} \, + \, e^{2V(r)} \, dr^{2} \, + \, e^{2W(r)}\,ds_{S^{3}}^{2} & , \\[2mm]
X & = & X(r) & , \\[2mm]
B_{(3)} & = & k(r)\,\textrm{vol}_{\textrm{Mkw}_{3}} \, + \, l(r)\,\textrm{vol}_{S^{3}} & , \\[2mm]
A^i\,&=\,&\frac{A(r)}{2\,g}\,\epsilon^{i\,k\,l}\,\omega_{j\,kl}\,d\,\theta^j& ,
\label{ansatz:mkw21vector}
\end{array}
\ee
where $S^3$ is parametrized by the parallelized basis $\{\theta^i\}$ introduced in \eqref{S3}.
As we mentioned in the previous section, we consider a Killing spinor $\zeta^a$ of the form \eqref{Kspinor} and satisfying \eqref{gamma3proj} and \eqref{vectorproj}. Thus $\zeta^a$ has two real independent components.
Plugging this \emph{Ansatz} into the Killing spinor equations \eqref{SUSY_eqns_7D}, we obtain the set of consistent first-order flow equations given in \eqref{flow:mkw21vector}. 
Remarkably, the coupling to 
the vector fields produces a set of consistent flow equations without any additional constraint as opposed to what happened in section \ref{sec:flows} for flows without vectors.

By solving the flow equations on 
a background of the form $\mathrm{Mkw}_{3}\times \mathbb{H}^3$ with a vanishing 4-form field strength and $A(r)=1$ we know that $\mathrm{AdS}_{4}\times \mathbb{H}^3$ solutions exist \cite{Acharya:2000mu} .
Thus it resoneable to wonder whether a solution of the same type with an $\mathrm{AdS}_{4}\times S^3$ background exists as a 
particular solution for \eqref{ansatz:AdS3vector}. However, one gets easily
convinced that such a solution cannot exist within the $\mathcal{N}=1$ truncation of the theory\footnote{Imposing $A(r)=1$, we found that the flow equations \eqref{flow:mkw21vector} and the equations
of motion \eqref{EOM_7D} are satisfied by a constant 3-form, by a linear dependence on $r$ of $U$ and by an imaginary constant value of $W$. }.
Moreover we observed that imposing $A(r)=1$ in \eqref{flow:mkw21vector} without any other specifications on the fields, the equations of motion do not admit any solutions. 

Thus we are forced to keep a non-trivial radial dependence for the gauge fields. In this case the flow equations \eqref{flow:mkw21vector} can be intregrated numerically. We are interested in those solutions that are
asymptotically locally $\mathrm{AdS}_7$, which means that we  first have to verify if there is a particular limit of the background in \eqref{ansatz:mkw21vector} reproducing $\mathrm{AdS}_7$ at the leading order in its asymptotic expansion.

\begin{figure}[htbp]
\centering
\vspace{0.1cm}
\includegraphics[width=55mm]{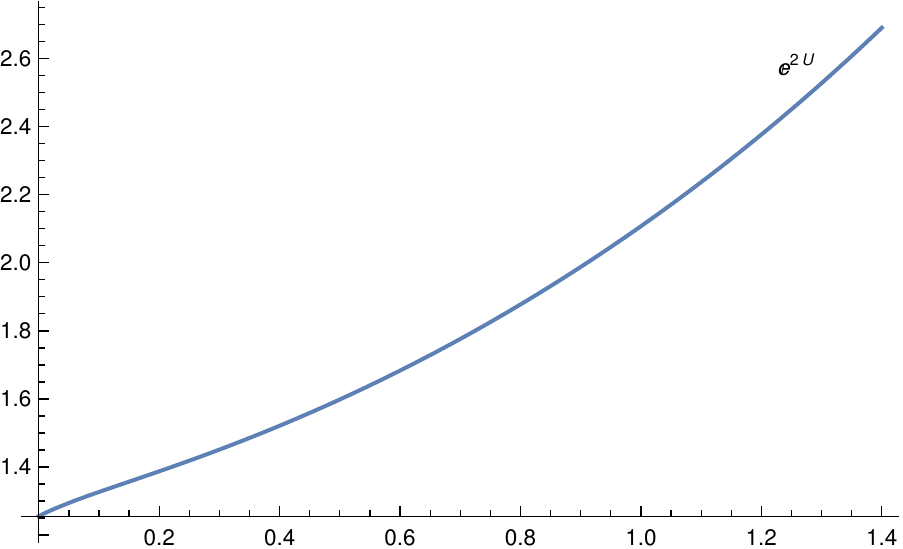}
\vspace{0.5cm}
\hspace{1.5cm}
\includegraphics[width=55mm]{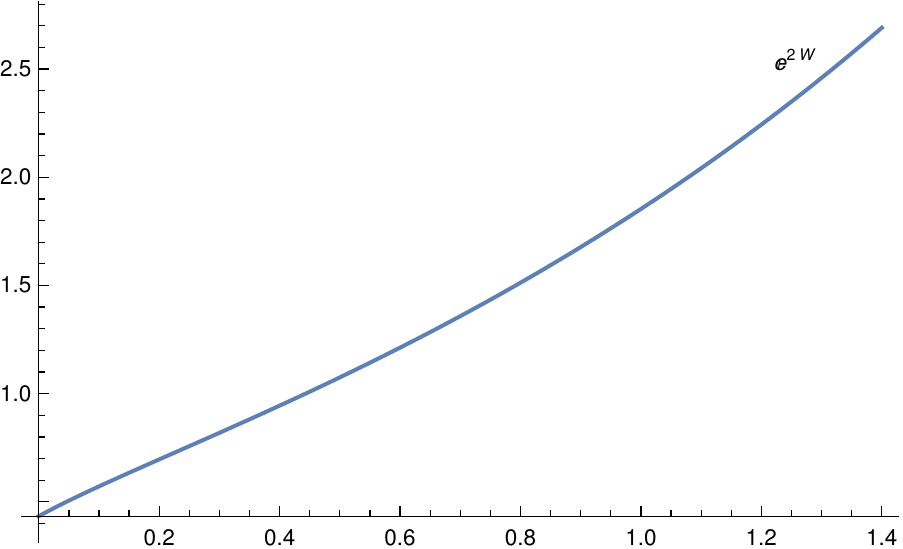}
\vspace{0.5cm}
\hspace{1.5cm}
\includegraphics[width=55mm]{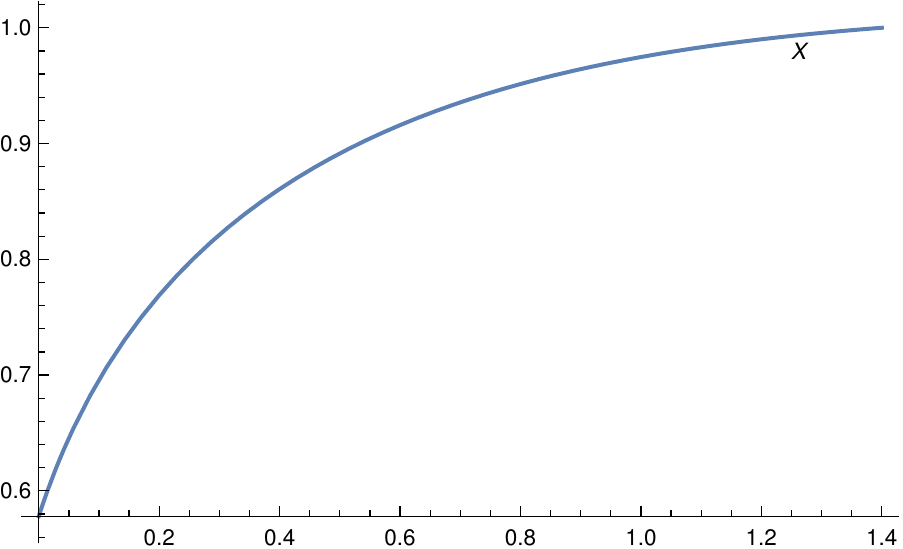}
\hspace{1.5cm}
\includegraphics[width=55mm]{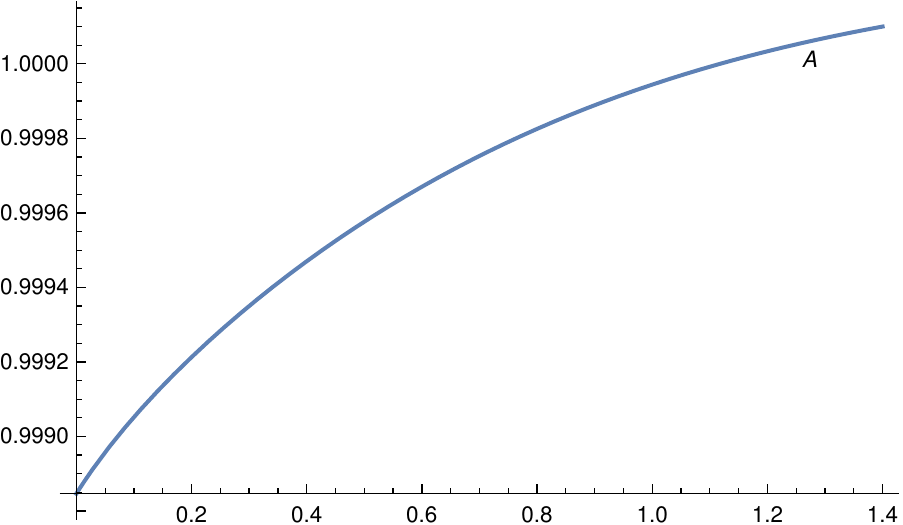}
\vspace{0.5cm}
\hspace{1.5cm}
\includegraphics[width=55mm]{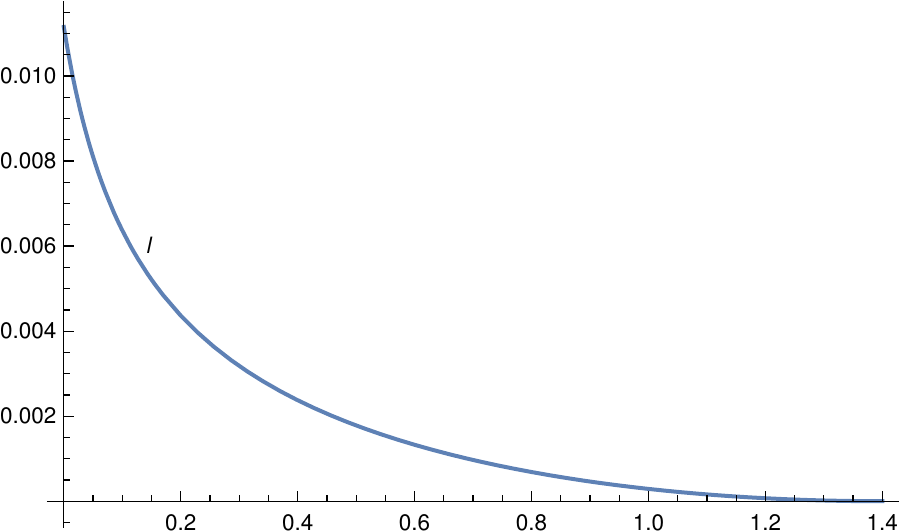}
\hspace{1.5cm}
\includegraphics[width=55mm]{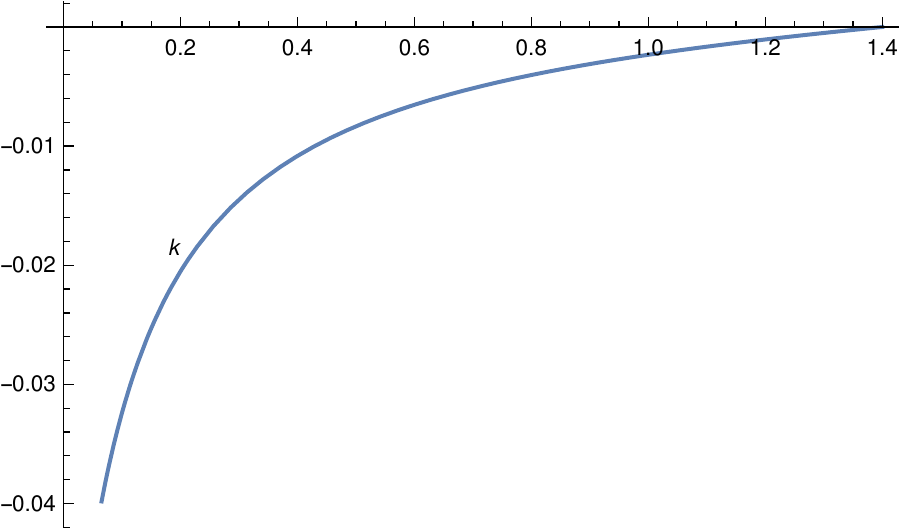}
\caption{\it Flows of $e^{2U(r)},\,e^{2W(r)},\,X(r),\,A(r),\,l(r),\,k(r)$ plotted in the interval $r\in (0,1.4]$ with $g=1$, $h=\frac{1}{2\sqrt{2}}$, $\kappa=1$ and $V(r)=0$.}
\label{mk21numerical}
\end{figure}

In order to be able to perform numerical integration, we also need to make a choice of the value of the free parameter in the system. In particular, we impose for simplicity $g=1$, $h=\frac{1}{2\sqrt{2}}$ and $\kappa=1$ and we make the gauge choice $V(r)=0$.
Then, it is possibile to verify that the following configuration
\be
\begin{split}
&U=\frac{r}{2\sqrt{2}}\ ,\quad W=\frac{r}{2\sqrt{2}}\ ,\quad X=1\ ,\quad \theta=0\  ,\quad Y=e^{U/2}\ ,\\
&k=0\ ,\quad \quad \quad l=0\ ,\quad \quad \quad A=1\ ,
\end{split}
\label{AdS7asymptotics}
\ee
solves \eqref{flow:mkw21vector} at the leading order when $r\rightarrow+\infty$. One can intregrate numerically \eqref{flow:mkw21vector}, by using the asymptotic behavior of the fields given in 
\eqref{AdS7asymptotics} as initial data. 
By doing so, one obtains a profile for the fields that is singular in $r\rightarrow 0$ and locally $\mathrm{AdS}_7$ for $r\rightarrow +\infty$. The explicit radial profile of the fields for this solution
is plotted in figure~\ref{mk21numerical}.

\subsection{Vectors coupled to the background $\mathrm{AdS}_3\times S^3$ }
\label{AdS3vector}

Let us now consider the background \eqref{backgroundAdS3} coupled to the vectors as given in \eqref{ansatzvector}. The complete \emph{Ansatz} is now given by
\be
\begin{array}{lcll}
ds_{7}^{2} & = & e^{2U(r)}\,ds_{\mathrm{AdS}_3}^{2} \, + \, e^{2V(r)} \, dr^{2} \, + \, e^{2W(r)}\,ds_{S^{3}}^{2} & , \\[2mm]
X & = & X(r) & , \\[2mm]
B_{(3)} & = & k(r)\,\textrm{vol}_{\mathrm{AdS}_3} \, + \, l(r)\,\textrm{vol}_{S^{3}} & , \\[2mm]
A^i\,&=\,&\frac{A(r)}{2\,g}\,\epsilon^{i\,k\,l}\,\omega_{j\,kl}\,d\,\theta^j\,,
\label{ansatz:AdS3vector}
\end{array}
\ee
where $\mathrm{AdS}_3$ and $S^3$ are respectively parametrized as in \eqref{AdS3} and \eqref{S3}. 

Given the Killing spinor $\zeta^a$ of the form \eqref{Kspinor} and satisfying \eqref{gamma3proj}, the set of the first-order flow equations describing the background \eqref{ansatz:AdS3vector} is
given in \eqref{flow:AdS3vector}.

The background \eqref{ansatz:AdS3vector} admits, among others, an $\mathrm{AdS}_3\times \mathbb{H}^4$ solution. This is an example of $\mathrm{AdS}_{d<7}$ solution with a non-constant
profile for both the gauge fields and the 3-form \cite{Gauntlett:2000ng}. In particular the following expressions for the fields,
\be
\begin{split}
&e^{2U}=\frac{L^2}{7}\ ,\quad \quad e^{2W}=\sinh\,(r)\ ,\quad \quad e^{2V}=1\ ,\quad \quad X=\left(\frac{7}{12}\right)^{1/5}\ ,\quad \quad \theta=\frac{\pi}{4}\ ,\quad \quad Y=e^{U/2}\ ,\\
&k=-\frac{2^{4/5}\, 3^{2/5}\, L^3}{ 7^{19/10}}\ ,\quad l=\frac{3^{2/5}}{2^{1/5}\, 7^{9/10}}\, \left(9 \cosh\, (r)-\cosh \, (3
   r)+8\right)\ ,\quad A=1+\cosh\,(r)\ ,
\end{split}
\label{AdS3H4}
\ee
provide a solution for both \eqref{flow:AdS3vector} and \eqref{EOM_7D} with $h=\frac{g}{2\sqrt{2}}$, $\kappa=2$ and $g=\frac{3^{1/5}\,7^{3/10}}{2^{1/10}}$.

\begin{figure}[htbp]
\centering
\vspace{0.1cm}
\includegraphics[width=55mm]{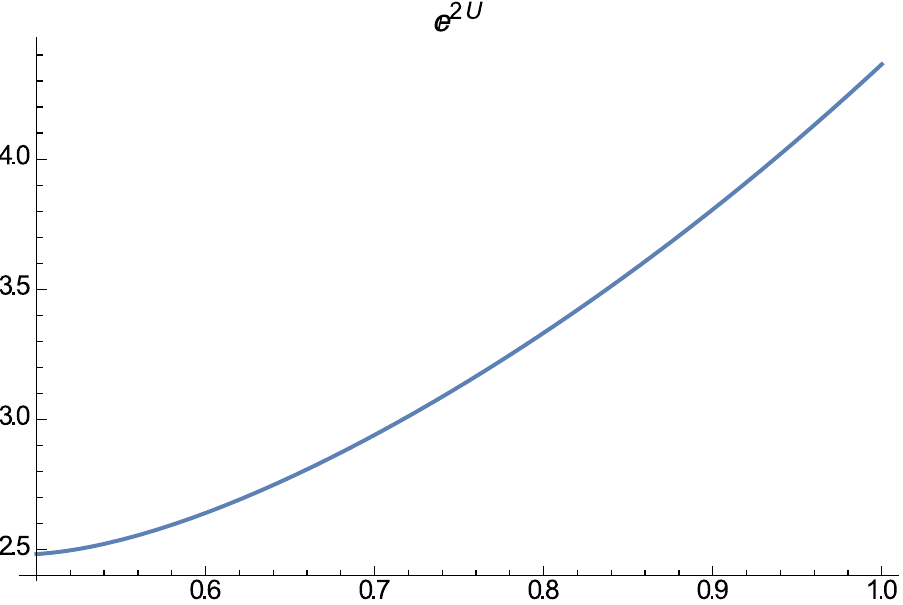}
\vspace{0.5cm}
\hspace{1.5cm}
\includegraphics[width=55mm]{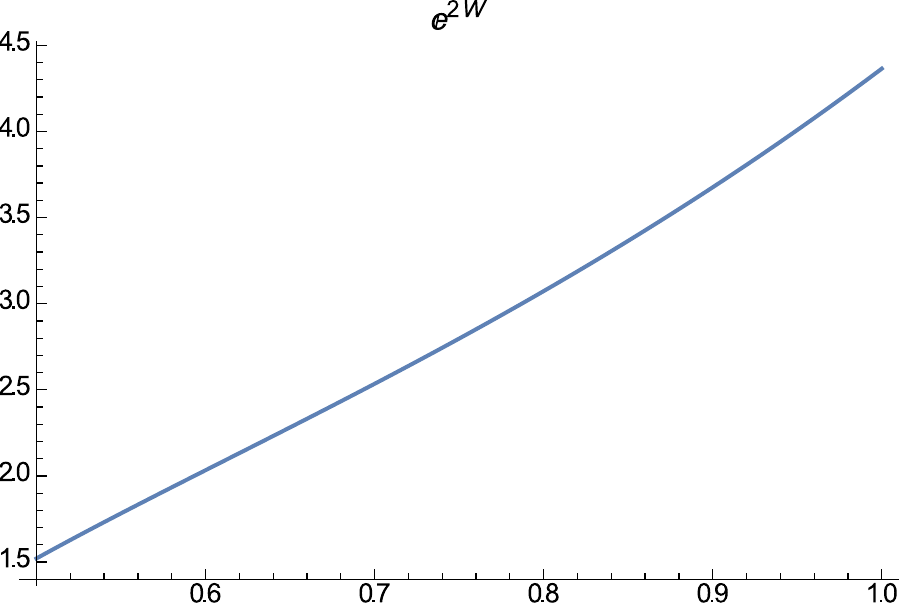}
\vspace{0.5cm}
\hspace{1.5cm}
\includegraphics[width=55mm]{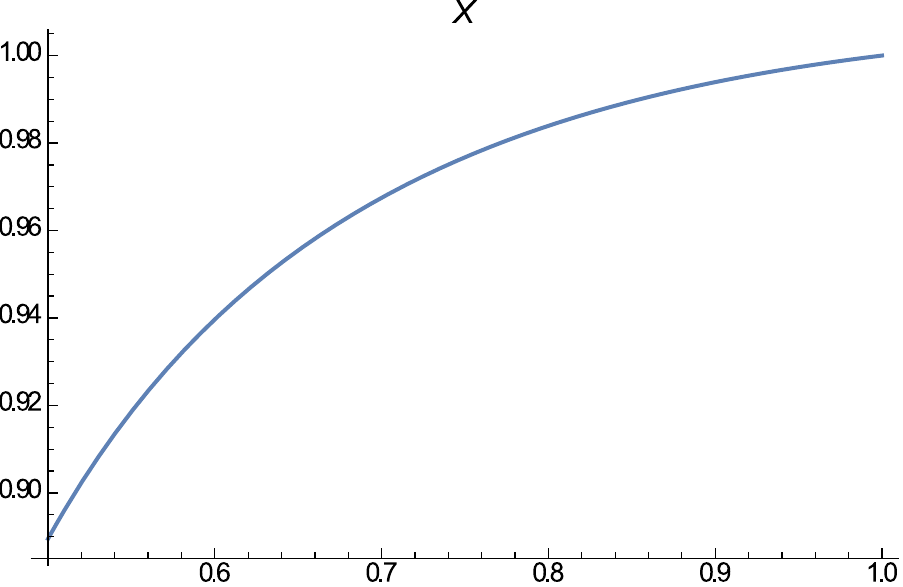}
\hspace{1.5cm}
\includegraphics[width=55mm]{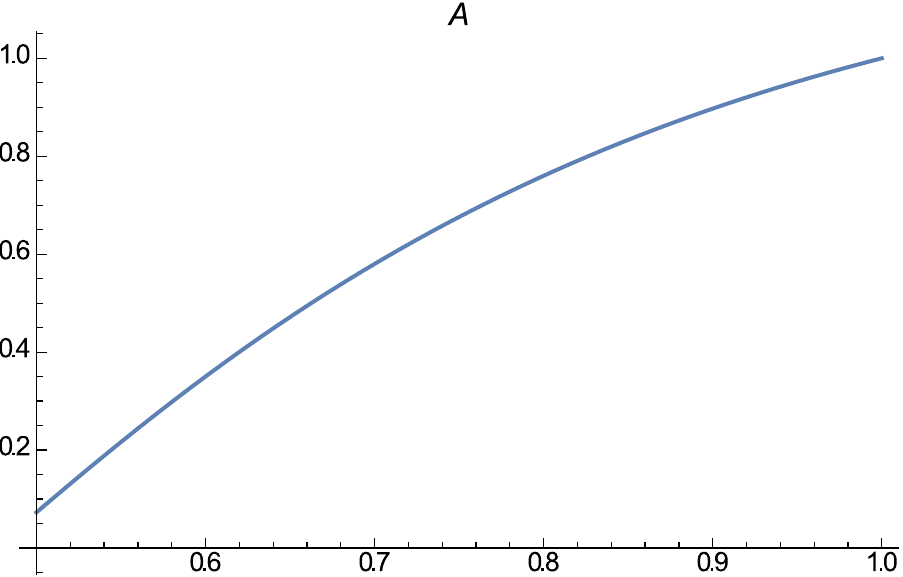}
\vspace{0.5cm}
\hspace{1.5cm}
\includegraphics[width=55mm]{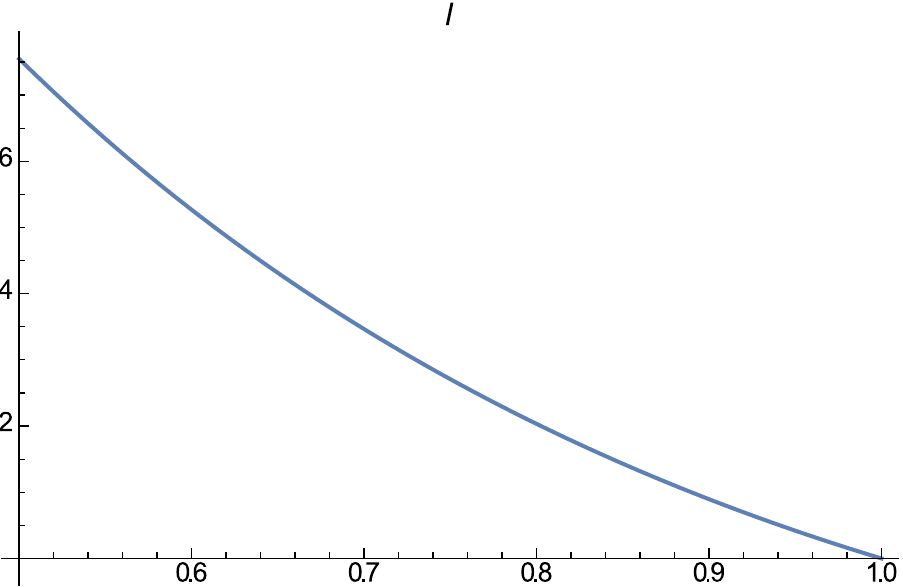}
\hspace{1.5cm}
\includegraphics[width=55mm]{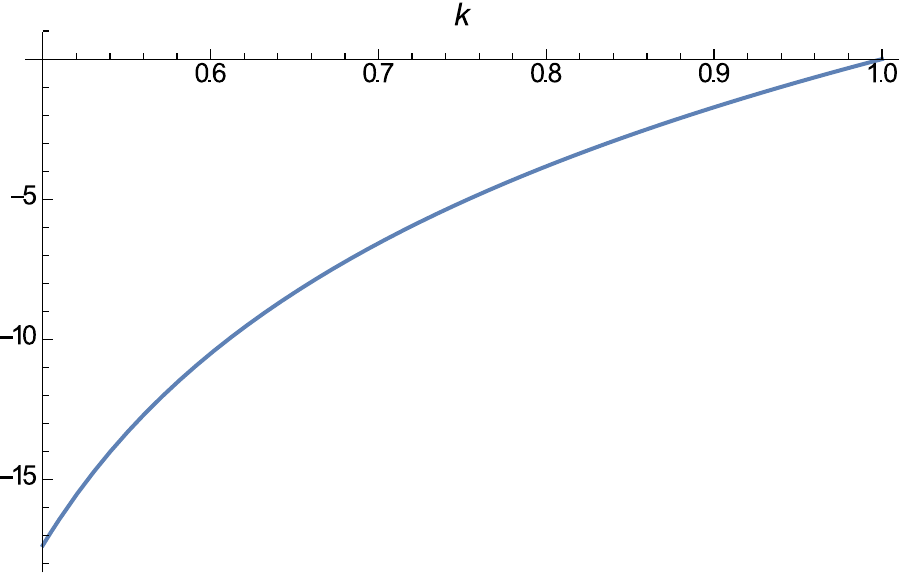}
\caption{\it Flows of $e^{2U(r)},\,e^{2W(r)},\,X(r),\,A(r),\,l(r),\,k(r)$ plotted in the interval $r\in (0.5,1]$ with $g=\frac{3^{1/5}\,7^{3/10}}{2^{1/10}}$, $h=\frac{g}{2\sqrt{2}}$, $\kappa=2$, $L=0.3$ and $V(r)=0$.}
\label{AdS3numerical}
\end{figure}

As in the previous section, we integrate numerically the flow equations \eqref{flow:AdS3vector} by starting from the locally $\mathrm{AdS}_7$ asymptotics. Choosing the same values for $g,\,h$ and $\kappa$ characterizing the 
solution \eqref{AdS3H4} and $V(r)=0$, it is possibile to show that the locally $\mathrm{AdS}_7$ configuration
\be
\begin{split}
&U=\frac{3^{1/5}\,7^{3/10}}{2^{8/5}}\,r\ ,\quad W=\frac{3^{1/5}\,7^{3/10}}{2^{8/5}}\,r\ ,\quad X=1\ ,\quad \theta=0\, ,\quad Y=e^{U/2}\ ,\\
&k=0\ ,\quad l=0\ ,\quad A=1\ ,
\end{split}
\label{AdS7asymptoticsAdS3}
\ee
solves \eqref{flow:AdS3vector} in the limit $r \rightarrow +\infty$.
We intregrated numerically the flow equations in \eqref{flow:AdS3vector} starting from \eqref{AdS7asymptoticsAdS3}. In this way we obtained
a flow that shows a singular behavior as $r\rightarrow 0$, while clearly keeping its locally $\mathrm{AdS}_7$ structure in its asymptotic region.
The explicit profile of the 7D fields is shown in figure~\ref{AdS3numerical}.
 It may be worth noticing that the above flow does not describe $\mathrm{AdS}_3\times \mathbb{H}^4$ in the limit where $r\rightarrow 0$, but this should not be a surprise since this solution describes
an $\mathrm{AdS}_3$ slicing of the 7D background, where the radial coordinate of the 7D background does not coincide with the radial coordinate of $\mathrm{AdS}_3$.
It is in fact this latter one which is expected to parametrize the flow where $\mathrm{AdS}_3$ emerges in the IR limit.
As for the complete flow realizing the full interpolation
between $\mathrm{AdS}_7$ and $\mathrm{AdS}_3\times \mathbb{H}^4$, it should be represented by a more general BPS background describing a $\mathrm{Mkw}_{2}$ slicing of the 7D background dependent on both coordinates.

%\newpage

\section{M-theory lifts}
\label{sec:Mtheory_lifts}

Given the solutions derived in section~\ref{sec:flows}, we will now try to give them an intepretation in terms of bound states in $\mathrm{M}$-theory. 
It is well known that the equations of motion of 
the minimal gauged supergravity in $D=7$ written in \eqref{EOM_7D} are obtained by reducing 11-dimensional supergravity on $S^4$ \cite{Lu:1999bc}. 
This consistent truncation produces the scalar potential \eqref{potential} depending on the parameters $g$ and $h$, where $h$ is related to the 11-dimensional $F_{(4)}$ flux and $g$ is the gauge parameters of the $\mathrm{\mathrm{SU(2)}}$ vectors describing 
the squashing of the 3-sphere with respect to which the $S^4$ is written as an $S^3$-fibration over a segment. 

This can be explicitly checked by means of a simple group-theoretical argument. To this end, we decompose the embedding tensor piece of the maximal theory with 11D origin from $S^{4}$, \emph{i.e.} the
$\textbf{15}$ of $\textrm{SL}(5,\mathbb{R})$, and identify the $\textrm{SO}(3)$-singlets corresponding to $h$ \& $g$. This procedure yields
\be
\begin{array}{ccccccc}
\textrm{SL}(5,\mathbb{R}) & \supset & \mathbb{R}^{+}_{1}\times\textrm{SL}(4,\mathbb{R}) & \supset & \mathbb{R}^{+}_{1}\times\mathbb{R}^{+}_{2}\times\textrm{SL}(3,\mathbb{R}) & \supset & \mathbb{R}^{+}_{1}\times\mathbb{R}^{+}_{2}\times\textrm{SO}(3) \\[3mm]
\textbf{15} & \rightarrow & \textbf{1}_{(+4)}\,\oplus\,\textbf{10}_{(-1)} & \rightarrow & \textbf{1}_{(+4;0)}\,\oplus\,\textbf{6}_{(-1;-2)} & \rightarrow & 
\underbrace{\textbf{1}_{(+4;0)}}_{h}\,\oplus\,\underbrace{\textbf{1}_{(-1;-2)}}_{g}
\end{array}\notag
\ee
Now one gets easily convinced that the 11D $F_{(4)}$ flux is to be identified with the embedding tensor piece which was already a singlet of $\textrm{SL}(4,\mathbb{R})$, \emph{i.e.} $h$, while the $\textrm{SU}(2)$
curvature is only expected to be a singlet of $\textrm{SO}(3)$, and hence is naturally identified with $g$.
 
Since the above gauging parameters are related to the 
fluxe configurations in the higher-dimensional theory, other reductions can be in principle considered and the simplest of those is certainly the one on the torus $\mathbb{T}^4$, yielding the potential \eqref{potential} with $g=0$.

 If one considers the flows obtained as solutions in $\mathcal{N}=1$ gauged supergravity in $D=7$, the existence of consistent truncations implies that
 the physics of some solitonic objects in $\mathrm{M}$-theory is captured by the solutions in 7-dimensional supergravity in the low-energy limit. 

The simplest example of this is given by the DW solutions in \eqref{DWflow} that describe three possible configurations in $\mathrm{M}$-theory, depending on how the gauging in the 7-dimensional supergravity is further specified. 
All of them consist of M5-branes reduced in different ways on their transverse space. In particular, one may easily see that \cite{Bergshoeff:2004nq}:
\begin{itemize}
 \item  $h\neq0$ and $g=0$: the DW \eqref{DWflow} describes an $\mathrm{M5}$-brane with four of its transverse coordinates reduced on a $\mathbb{T}^4$, 
  \item $h=0$ and $g\neq0$: the DW \eqref{DWflow} describes an $\mathrm{NS5}$-brane in $\mathrm{IIA}$ string theory reduced on an $S^3$, 
or an $\mathrm{M5}$-brane with four of its transverse coordinates reduced on $S^1\times S^3$,
   \item  $h\neq0$ and $g\neq 0$: the DW \eqref{DWflow} describes an $\mathrm{M5}$-brane with four of its transverse coordinates reduced on an $S^4$.
 \end{itemize}

As a general fact, not all of the truncations of higher-dimensional theories admit solutions with an $\mathrm{AdS}_7$ asymptotic behavior. 
In fact, since only the complete form of the potential \eqref{potential} admits $\mathrm{AdS}_7$ critical points,
 only the last DW solution (with $h\neq0$ and $g\neq 0$) will asymptote to the $\mathrm{AdS}_7$ that is associated with the $\mathrm{AdS}_7\times S^4$ Freund-Rubin vacuum. 
Such a vacuum can be indeed obtained by taking the near-horizon limit of the $\mathrm{M5}$-brane geometry.

\subsection{Dyonic solutions and $\mathrm{M2-M5}$ bound state on $\mathbb{T}^4$}

Moving to the flows involving a dyonic profile for the 3-form potential, let us first consider the solution presented in \eqref{ch_DW_noAdS}. In this case the potential driving the solution is given by
\be
\mathcal{V}(X) \ = \ 2h^{2} \, X^{-8}\ ,
\label{runawaypotential}
\ee
which, due to its run-away behavior, has no critical points. As we said, the truncation producing a potential with $g= 0$ is obtained by considering the low-energy limit of $\mathrm{M}$-theory on a 4-torus 
$\mathbb{T}^4$ with non-vanishing 4-form flux. 
In this section we want to show that the flow in \eqref{ch_DW_noAdS} is the low-energy description of a supersymmetric 
$\mathrm{M2-M5}$ bound state discovered in \cite{Izquierdo:1995ms} by uplifting to eleven dimensions a dyonic membrane solution obtained in 
$\mathcal{N}=2$, $D=8$ supergravity.

The corresponding eleven-dimensional background reads
\be
\begin{split}
 ds^2_{11}&\,=\,H^{-2/3}\,\left(\sin^2 \xi +H \, \cos^2\xi \right)^{1/3}\,ds^2_{\textrm{Mkw}_{3}}+
 H^{1/3}\,\left(\sin^2 \xi +H \, \cos^2\xi \right)^{1/3}\,ds^2_{\mathbb{R}^5}\\
 &\ +\,H^{1/3}\,\left(\sin^2 \xi +H \, \cos^2\xi \right)^{-2/3}\,ds^2_{\mathbb{R}^3}\ ,
\end{split}
\label{lambert:metric}
\ee
where $H$ is a harmonic function on $\mathbb{R}^5$ and $\xi$ is a constant angle. The 4-form field strength is given by
\be
F_{(4)}\,=\,\frac12 \,\cos \xi \, *_{(5)} d\,H+\frac12\,\sin \xi \, d\,H^{-1} \wedge \mathrm{vol}_{\textrm{Mkw}_{3}}-\frac{3\,\sin(2\,\xi)}{2\,\left(\sin^2 \xi +H \, \cos^2\xi \right)^{2}}\,\mathrm{vol}_{\mathbb{R}^3}\,\wedge dH\ ,
\label{lambert:F11}
\ee
where $\mathrm{vol}_{\textrm{Mkw}_{3}}$ and $\mathrm{vol}_{\mathbb{R}^3}$ are respectively the volume of the 3-dimensional Minkowski space and the volume of $\mathbb{R}^3$.

Since $H$ is defined on $\mathbb{R}^5$, the solution may be interpreted as the effective 
description of an $\mathrm{M2}$-brane completely smeared over the worldvolume of an $\mathrm{M5}$ or, equivalently, of an $\mathrm{M5}$-brane
carrying a dissolved $\mathrm{M2}$ charge. This configuration preserves $16$ supercharges. Note that it is not the mere
superposition between the $\mathrm{M2}$ and the $\mathrm{M5}$-brane and this is due to the presence
of the third term of \eqref{lambert:F11} accounting for M2 -- M5 interactions.
There are two particular values for the parameter $\xi$:
\begin{itemize}
 \item $\cos \xi=0$: purely electric case corresponding to a pure (smeared) $\mathrm{M2}$-brane,
  \item $\sin \xi=0$: purely magnetic case corresponding to a pure $\mathrm{M5}$-brane.
\end{itemize}
Because of the intrinsic structure of bound state of the solution \eqref{lambert:metric}, its brane
interpretation for general values of $\xi$ remains somewhat obscure\footnote{This issue was originally discussed in \cite{Niarchos:2012pn,Niarchos:2013ia}, where this 11D solution at generic angles $\xi$ 
was given an interpretation in terms of an M2 -- M5 funnel geometry.}, 
but it can be shown that it has a smooth horizon for any $\xi\neq\frac{\pi}{2}$, the corresponding near-horizon geometry
being $\mathrm{AdS}_7\times S^4$. 

From \cite{Izquierdo:1995ms} we know that, by compactifying\footnote{Giving a periodic identification on the coordinates of $\mathbb{R}^3$.} \eqref{lambert:metric} on a $\mathbb{T}^3$, one obtains a flow in $\mathcal{N}=2$, $D=8$ supergravity featured by a dyonic
3-form and an axio-dilaton. The 8-dimensional flow trasforms under $\textrm{SL}(2,\mathbb{R})$ and this means
that one can always find a transformation such that the 8-dimensional 3-form is completely electric.

Let us now reduce the \eqref{lambert:metric} and \eqref{lambert:F11} on a $\mathbb{T}^4$.
In order for this procedure to be consistent, a \emph{smearing} of the charge distrubution is required along all the $\mathbb{T}^4$ coordinates. To implement this, choose the $\mathbb{R}^5$ coordinates such that
\be
ds^2_{\mathbb{R}^5}\,=\,dz^2+ds^2_{\mathbb{T}^4}\,,\qquad \text{with} \qquad H=1+\alpha \, z\ ,
\ee
with $\alpha$ real parameter. The form of \eqref{lambert:F11} suggests a dyonic profile for the corresponding 7-dimensional 3-form, but in this case the odd-dimensional self-duality conditions \eqref{SD_cond} spoil the possibility of rotating the dyonic 3-form
into a completely electric one as it was done in the 8-dimensional case.

The reduction on $\mathbb{T}^4$ of 11-dimensional supergravity can be performed directly at the level of the 
11-dimensional action
 with the following reduction \emph{Ansatz} on the metric,
\be
ds_{11}^2\,=\,X^{-4/3}\, ds^2_{7}\,+\, X^{5/3} \,ds^2_{\mathbb{T}^4}\ ,
\label{reductedmetric}
\ee
and including a 4-form field strength wrapping the $\mathbb{T}^4$,
\be
F_{(4)}=q\,\mathrm{vol}_{\mathbb{T}^4}\ ,
\label{G4}
\ee
where $q$ is the flux associated to the 11-dimensional 3-form and $X$ is
the scalar field belonging to the supergravity multiplet of the 7-dimensional minimal supergravity associated to the volume modulus of $\mathbb{T}^4$. Imposing\footnote{We imposed the relation
 $\frac{\mathrm{vol}_{\mathbb{T}^4}}{2\,\kappa_{11}^2}\equiv\frac{1}{2\,\kappa_7^2}$ between the gravitational couplings.}  this reduction \emph{Ansatz} we obtain the action \eqref{Lagrangian_7D} with $g=0$, $A^i=0$ and
a scalar potential given by \eqref{runawaypotential}.

Using the reduction \emph{Ansatz} \eqref{reductedmetric} and \eqref{G4}, we want to compare \eqref{lambert:metric} and \eqref{lambert:F11} with \eqref{ch_DW_noAdS}. We start by extracting the 7-dimensional flow from 
\eqref{lambert:metric} and \eqref{lambert:F11}.

Let us begin with
the first term of \eqref{lambert:F11} placed on $\mathbb{T}^4$, \emph{i.e.} $*_{(5)} d\,H\,=\,\alpha\,\mathrm{vol}_{\mathbb{T}^4}$.
Comparing it with \eqref{G4}, we immediately obtain $q=\frac{\cos \xi}{2}\,\alpha$.
 By a comparison with \eqref{reductedmetric}, it is possibile to extract a 7-dimensional metric and the
expression for $X$ from \eqref{lambert:metric}, one obtains
\be
\begin{split}
ds_7^2&\,=\,  H^{-2/5}\,\left(\sin^2 \xi +H \, \cos^2\xi \right)^{3/5}\,ds^2_{\textrm{Mkw}_{3}}+
 H^{3/5}\,\left(\sin^2 \xi +H \, \cos^2\xi \right)^{3/5}\,dz^2\\
 &\ +\,H^{3/5}\,\left(\sin^2 \xi +H \, \cos^2\xi \right)^{-2/5}\,ds^2_{\mathbb{R}^3}\ ,\\
 X&\,=\, H^{1/5}\,\left(\sin^2 \xi +H \, \cos^2\xi \right)^{1/5}\ .
 \end{split}
 \label{lambert7D:metric}
\ee
The 7-dimensional 4-form field strength is simply given by the second and the third terms of \eqref{lambert:F11},
in particular one has
\be
\mathcal{F}_{(4)}\,=\,\frac{\sin\xi}{2}\,d\, \left( H^{-1}\,\mathrm{vol}_{\textrm{Mkw}_{3}} \right)+\frac{3\,\sin\xi}{\cos \xi}
\,d\, \left(\left(\sin^2 \xi +H \, \cos^2\xi \right)^{-1}\,\mathrm{vol}_{\mathbb{R}^3} \right)\ .
 \label{lambert7D:4form}
\ee
We can now consider the flow \eqref{ch_DW_noAdS} and compare it with \eqref{lambert7D:metric} and \eqref{lambert7D:4form}.
We firstly rewrite \eqref{ch_DW_noAdS} with a more general dependence on the integration constants that will be
fixed by the matching,
\be
\label{ch_DW_noAdS_constants}
\begin{array}{lclclc}
e^{2U} \, = 2^{1/5}\,e^{2\alpha_1} \sinh(2hr)^{1/5}\,\coth(hr) & & , & & e^{V} \, = 2^{8/5}\,\alpha_3^4 \sinh(2hr)^{8/5} & , \\[2mm]
e^{2W} \, = 2^{1/5}\,e^{2\alpha_2}\, \sinh(2hr)^{1/5}\,\tanh(hr) & & , & & X \, = 2^{2/5}\,\alpha_3 \sinh(2hr)^{2/5} & , \\[2mm]
k \, = \,\frac{e^{3\alpha_1}}{2\,\alpha_3^2} \,\sinh(hr)^{-2} & & , & & l \, = -\frac{e^{3\alpha_2}}{2\,\alpha_3^2}\, \cosh(hr)^{-2} & . \\[2mm]
\end{array}
\ee
One finds that for the following values of the constants $\alpha_1,\,\alpha_2,\,\alpha_3$ and of $H$,
\be
\begin{split}
 &e^{\alpha_1}\,=\,2^{-1/5}\cos(\xi)^{2/5}\sin(\xi)^{1/5}\ ,\\
 &e^{\alpha_2}\,=\,-2^{-1/5}\cos(\xi)^{-3/5}\sin(\xi)^{1/5}\ ,\\
 &\alpha_3^2\,=\,2^{-8/5}\cos(\xi)^{-4/5}\sin(\xi)^{8/5}\ ,\\
 &H\,=\,2\,\sin\xi \, \alpha_3^2 \,e^{-3\alpha_1}\,\sinh(hr)^{2}\ .
 \label{constants}
\end{split}
\ee
the functions $U,\,W,\,X,\,l,\,k$ describing the flow \eqref{ch_DW_noAdS_constants} match exactly with \eqref{lambert7D:metric} and \eqref{lambert7D:4form}.
We can finally derive the relation between the 7-dimensonal radial coordinate $r$ and
the radial coordinate of $\mathrm{M}$-theory $z$ by comparing the the radial parts of the 7-dimensional metrics,
\be
e^V\,dr\overset{!}{=}H^{3/10}\,\left(\sin^2 \xi +H \, \cos^2\xi \right)^{3/10}\,dz\ .
\label{eqzr}
\ee
Using \eqref{constants} and integrating \eqref{eqzr} we obtain
\be
z\,=\,\frac{\sin^2\xi}{4\,h\,\cos\xi}\,\cosh(2hr)\,+\,z_0\ .
\label{Mradius}
\ee
The constant $z_0$ can be determined by comparing $H=1+\alpha\,z$ with the expression of $H$ written in \eqref{constants} obtaining
\be
\alpha=\frac{2\,h}{\cos \xi}\,, \qquad z_0=-\frac{1+\cos^2\xi}{4h\cos\xi}\ .
\ee
Recalling that $q=\frac{\cos \xi}{2}\,\alpha$, one finds
\be
q\,=\,h\ .
\ee
We conclude that the flow \eqref{ch_DW_noAdS_constants} obtained in minimal supergravity in $D=7$ and described
by a dyonic 3-form and by the potential \eqref{runawaypotential} describes the low-energy limit of 
the $\mathrm{M2}-\mathrm{M5}$ reduced on $\mathbb{T}^4$. In particular the St{\"u}ckelberg mass $h$ is identified with the flux associated to 
the 11-dimensional 4-form field strength wrapping the 4-torus.

\subsection{$\mathrm{AdS}_7$ flows and $S^4$ reductions}

Let us now move to considering the asymptotically $\mathrm{AdS}_7$ flows derived in section \ref{sec:flows}
and their $\mathrm{M}$-theory picture. The main difference with respect to the case
of \eqref{ch_DW_noAdS} is the $\mathrm{AdS}_7$ asymptotic behavior that extremizes the potential
\be
\mathcal{V}(X) \ = \ 2h^{2} \, X^{-8} \, - \, 4\sqrt{2}\,gh\, X^{-3} \, - \, 2g^{2} \,X^{2} \  .
\label{pot}
\ee
The truncation of 11-dimensional supergravity describing \eqref{pot} is the one on a squashed $S^4$ \cite{Lu:1999bc} and 
it is defined by the complete $\mathcal{N}=1$, $D=7$ supergravity multiplet $(g_{\mu\nu},\,X,\,B_{(3)},\,A^i)$ whose
equations of motion, supplemented with the odd-dimensional self-duality conditions are written in \eqref{EOM_7D} and \eqref{SD_cond}. 

The metric of the internal $S^4$ is given by a foliation
of 3-spheres and its deformations are parametrized by the 7-dimensional scalar $X$. The squashing
leaves the 3-sphere foliations preserved. Thus,
introducing the basis of left-invariant forms $\eta^i$ on the 3-sphere, the 7-dimensional gauge fields $A^i$
describe the $\mathrm{\mathrm{\mathrm{SU(2)}}}$ bundle over the $S^3$ and the metric of the internal space is given by
\be
  ds_4^2 \,=\,X^3\,\Delta \, d\psi ^2\,+\,\frac{X^{-1}}{4}\,\cos^2\psi \sum_{i=1}^3\,(\eta^i-g\,A^i)^{2} \ , \quad \text{with}\quad 
 \Delta\,\equiv \,X^{-4}\sin^2\psi+X\,\cos^2\psi\ .
 \label{S^4squashed}
\ee
The truncation holds at the level of the equations of
motion and of the odd-dimensional self-duality conditions \eqref{EOM_7D} \& \eqref{SD_cond}, and it is specified by the following 11-dimensional \emph{Ansatz},
\be
\begin{split}
 ds_{11}^2&\,=\,\Delta^{1/3}\,ds^2_7+2\,g^{-2}\,\Delta^{-2/3}\,ds_4^2\ ,\\
 A_{(3)}&\,=\,\sin\psi \, B_{(3)}+\frac{g^{-3}}{2\sqrt{2}}\,\left(2\,\sin\psi+\sin\psi\,\cos^2\psi\,\Delta^{-1}\,X^{-4} \right)\,\epsilon_{(3)}\\
 &-\frac{g^{-2}}{\sqrt{2}}\,\sin\psi \,F^i_{(2)}\wedge h^i-\frac{g^{-1}}{\sqrt{2}}\,\sin\psi \, \omega_{(3)}\ ,
 \label{S^4reduction}
\end{split}
\ee
where $h^i\,\equiv\,\eta^i-g\,A^i$, $\epsilon_{(3)}\,\equiv\,h^1\wedge h^2 \wedge h^3$ and 
$\omega_{(3)}\,\equiv\,A^i\wedge \mathcal{F}_{(2)}^i-\frac16\,g\,\epsilon_{ijk}\,A^i\wedge A^j\wedge A^k$, and the fields
$X,\,B_{(3)}$ and $A^i$ are functions of the 7-dimensional background.

The flows with an $\mathrm{AdS}_7$ asymptotic behavior obtained in section \ref{sec:flows} can be organized in the following two groups:
\begin{itemize}
 \item $\mathrm{Mkw}_{3}\times S^3$ backgrounds\ ,
 \item $\mathrm{AdS}_3\times S^3$ backgrounds\ .
\end{itemize}
Furthermore in both cases we presented flows with and without the coupling to $\mathrm{\mathrm{\mathrm{SU(2)}}}$ vectors. By means of the uplift formula in \eqref{S^4reduction}, it is possibile to lift the 7-dimensional flows
given in \eqref{ch_DW_AdS}, \eqref{AdS3_DW_AdS_U=V}, \eqref{AdS3_DW_AdS} to eleven dimensions, while the existence of numerical flows obtained by solving \eqref{flow:mkw21vector} and \eqref{flow:AdS3vector} ensures the existence of 
corresponding 11-dimensional configurations.

We know that all the $\mathrm{AdS}_7$
flows of section \ref{sec:flows} are described by a dyonic profile for the 3-form that cannot be recast into a purely electric form because of the odd-dimensional self-duality conditions in \eqref{SD_cond}. 
Due to this argument, we are then again forced into considering
 $\mathrm{M2}-\mathrm{M5}$ bound states described in 11-dimensional supergravity by the solution \eqref{lambert:metric} and \eqref{lambert:F11}. 
This solution has an $\mathrm{AdS}_7\times S^4$ near-horizon geometry compatible
with the asymptotics of our 7-dimensional flows and a dyonic profile of the 3-form once compactified on $S^4$, but the issue here is to find a suitable coordinate system for the uplifted solutions such that a clean brane picture arises.
This is particularly manifest for the flow \eqref{ch_DW_AdS} 
coming from the $\mathrm{Mkw}_{3}\times S^3$ where such diffeomorphisms on the uplifted flow should relate the coordinates $(r,\,\psi)$ with the radial coordinate of $\mathrm{M}$-theory. 

Giving an interpretation of the warped solutions
\eqref{AdS3_DW_AdS_U=V} and \eqref{AdS3_DW_AdS} is more difficult since the presence of the $\mathrm{AdS}_3$ slicing implies a modification of the brane picture. For example, the semi-localized intersection of a pp-wave with an $\mathrm{M5}$-brane 
would modify the geometry of the worldvolume of the $\mathrm{M5}$ producing $\mathrm{AdS}_3$ in the near-horizon limit \cite{Cvetic:2000cj}. 
This may in principle hold true even when constructing an intersection of the $\mathrm{M2}-\mathrm{M5}$ bound state with a pp-wave, but it is in general difficult to specify the concrete momentum charge
distribution realizing it.

Finally the flows involving vectors should describe the wrapping of the worldvolume of the $\mathrm{M2}-\mathrm{M5}$ bound state on an $S^3$. However, since in this case we are even lacking the analytic form of the flows, 
it becomes technically impossible to search for the correct coordinate system which could verify our expectations. 
On the hand of course, the presence of the twisting condition \eqref{twisting} guaranteeing some residual supersymmetry suggests some spontaneous brane wrapping mechanism.

\section{Conclusions}
\label{sec:conclusions}

In this paper we considered minimal gauged supergravity in seven dimensions with $\textrm{SU}(2)$ gauge group and non-vanishing topological mass. The field content of the supergravity multiplet is given by the graviton, a scalar field $X$, a 
3-form $B_{(3)}$ and three $\mathrm{\mathrm{\mathrm{SU(2)}}}$ vector fields. We presented various novel solutions in this theory with backgrounds defined by a $\mathrm{Mkw}_{3}$ and $\mathrm{AdS}_3$ slicing. In the absence of vectors the first-order flow equations are solved
analitically, while only numerically when vectors are coupled. In particular we found a few examples of asymptotically locally $\mathrm{AdS}_{7}$ flows with a non-trivial profile of the 3-form $B_{(3)}$.

Particularly intersting are the flows describing an $\mathrm{AdS}_3$ slicing of the 7-dimensional spacetime. The brane picture in $\mathrm{M}$-theory of these solutions is not clear and its understanding could be especially 
relevant for their $\mathrm{AdS}/\mathrm{CFT}$ applications, since the holografic interpretation of these warped-$\mathrm{AdS}_3$ flows should be related to a conformal defect in the $(1,0)$ $\mathrm{SCFT}$ in $D=6$, in the spirit of 
\cite{Lunin:2007ab}. 
Furthermore these warped solutions should imply the existence of a new class
of RG flows across dimensions between the $(1,0)$ $\mathrm{SCFT}$ and a $\mathrm{SCFT}$ in $D=2$. 
Such flows are expected to be described by a $\mathrm{Mkw_{2}}$ slicing of spacetime depending two coordinates (the radial coordinate of $\mathrm{AdS}_3$ and the one of the 7-dimensional background).
Finally, the warped structure of the flows presented here suggests the possibility of studying truncations of minimal gauged supergravity in $D=7$ to a gauged supergravity in $D=3$ and this could be of 
great interest also in relation to a classification of $\mathrm{AdS}_3$ solutions of massive type IIA supergravity.
We hope to come back to these points in the future.

%%%%%%%%%%%%%%%%%%%%%%%%%%%%%%%%%%%%
%
% Acknowledgments
%
%%%%%%%%%%%%%%%%%%%%%%%%%%%%%%%%%%%%

\section*{Acknowledgments}

The work of GD is supported by the Swedish Research Council (VR), and the G\"oran Gustafsson Foundation. NP would like to thank Dietmar Klemm and Marco Rabbiosi
for support and useful discussions, and the members of the Department of Theoretical Physics at the Uppsala University for their kind hospitality while this work was being prepared.

\newpage

%%%%%%%%%%%%%%%%%%%%%%%%%%%%%%%%%%%%
%
% Appendices
%
%%%%%%%%%%%%%%%%%%%%%%%%%%%%%%%%%%%%

\appendix

\section{Symplectic-Majorana spinors in $D=7$}
\label{app:SM_spinors}

In this appendix we summarize the set of relevant conventions concerning irreducible spinors in $1+6$ dimensions and the corresponding representation of the Dirac matrices which we adopt throughout this
work. In $D=7$ with Lorentzian signature, the irreducible spinors are of Dirac type and carry $2^{\textrm{\textlbrackdbl} 7/2 \textrm{\textrbrackdbl}}\,=\,8$ complex components. 
The same degrees of freedom may be then rearranged into a symplectic-Majorana (SM) spinor, \emph{i.e.} an $\textrm{SU}(2)_{R}$ doublet of spinors $\zeta^{a}$ satisfying a \emph{pseudo-reality condition}
of the form
\be
\label{SM_cond}
\zeta_{a}\ \equiv \ \left(\zeta^{a}\right)^{*} \ \overset{!}{=} \ \epsilon_{ab}\,B\,\zeta^{b} \ ,
\ee
where $\epsilon_{ab}$ denotes the $\textrm{SU}(2)$-invariant Levi-Civita symbol, and $B$ is the matrix that controls complex conjugation of Dirac spinors. 
Note that the condition \eqref{SM_cond} makes sure that the amount of on-shell real degrees of freedom described by $\zeta$ be $16$.
The Dirac matrices $\left\{\gamma^{m}\right\}_{m\,=\,0,\,\cdots\,6}$ satisfy
\be
\left\{\gamma^{m},\,\gamma^{n}\right\} \ = \ 2\,\eta^{mn} \, \mathds{1}_{8} \ ,
\ee
where $\eta\,\equiv\,\textrm{diag}(-1,+1,+1,+1,+1,+1,+1)$.

We adopted the following explicit representation for the Clifford algebra \cite{VanProeyen:1999ni}
\be
\label{rep_gamma}
\begin{array}{lclc}
\gamma^{0} & = & i \, \sigma^{2} \, \otimes \, \mathds{1}_{2} \, \otimes \, \mathds{1}_{2} & , \\[1mm]
\gamma^{1} & = & \,\, \sigma^{1} \, \otimes \, \mathds{1}_{2} \, \otimes \, \mathds{1}_{2} & , \\[1mm]
\gamma^{2} & = & \,\, \sigma^{3} \, \otimes \, \sigma^{1} \, \otimes \, \mathds{1}_{2} & , \\[1mm]
\gamma^{3} & = & \,\, \sigma^{3} \, \otimes \, \sigma^{3} \, \otimes \, \mathds{1}_{2} & , \\[1mm]
\gamma^{4} & = & \,\, \sigma^{3} \, \otimes \, \sigma^{2} \, \otimes \, \sigma^{1} & , \\[1mm]
\gamma^{5} & = & \,\, \sigma^{3} \, \otimes \, \sigma^{2} \, \otimes \, \sigma^{2} & , \\[1mm]
\gamma^{6} & = & \,\, \sigma^{3} \, \otimes \, \sigma^{2} \, \otimes \, \sigma^{3} & , 
\end{array}
\ee
where $\left\{\sigma^{i}\right\}_{i\,=\,1,\,2,\,3}$ are the Pauli matrices
\be
\label{Pauli}
\s^{1} =
\left(
\begin{array}{cc}
0 & 1  \\
1 & 0
\end{array}
\right)
\hspace{5mm} \textrm{ , } \hspace{5mm}
\s^{2} =
\left(
\begin{array}{cc}
0 & -i  \\
i & 0
\end{array}
\right)
\hspace{5mm} \textrm{ , } \hspace{5mm}
\s^{3} =
\left(
\begin{array}{cc}
1 & 0  \\
0 & -1
\end{array}
\right) \ .
\ee
One can check that the representation given in \eqref{rep_gamma} satisfies the following identity
\be
\g_{*} \ \equiv \ \g^{0}\,\g^{1}\,\g^{2}\,\g^{3}\,\g^{4}\,\g^{5}\,\g^{6} \ = \ \mathds{1}_{8} \ .
\ee
In this spacetime signature the matrices $A$, $B$ and $C$ which respectively realize Dirac, complex and charge conjugation of spinors, have the following defining properties
\be
\begin{array}{lclclclclclc}
\left(\g^{m}\right)^{\dagger} & = & -A \, \g^{m} \, A^{-1} & , & \left(\g^{m}\right)^{*} & = & B \, \g^{m} \, B^{-1} & , & \left(\g^{m}\right)^{T} & = & -C \, \g^{m} \, C^{-1} & . 
\end{array}
\ee
One can check that a consistent choice for the above operators w.r.t. our representation given in \eqref{rep_gamma} is given by
\be
\begin{array}{lclclc}
A \ = \ \g^{0} & , & B \ = \ -i \, \g^{46}  & , & C \ = \ i \, \g^{046} & , 
\end{array}
\ee
which satisfy
\be
\begin{array}{lclclc}
B^{T} \ = \ C \, A^{-1} & , & B^{*}\,B \ = \ -\mathds{1}_{8}  & , & C^{T} \ = \ -C^{-1} \ = \ -C^{\dagger} \ = \ C & .
\end{array}
\ee
\newpage
\section{Flow equations in presence of vectors: $\mathrm{Mkw}_{3}\times S^3$}
\label{floweq:mk21_S3_vector}

In what follows the first-order flow equations that have been solved numerically in section \ref{mkw21vector} are given.
Let's consider the \emph{Ansatz} \eqref{ansatz:mkw21vector} with Killing spinor $\zeta^a$ of the form \eqref{Kspinor} and satisfying \eqref{gamma3proj} and \eqref{vectorproj}.
From \eqref{SUSY_eqns_7D} we obtain the following system of first-order flow equations,
\be
\begin{split}
U'\,=\,&\frac{ e^{V-2 W}}{80\, g\, X\,\cos (2 \theta )}\,\biggl[ 6\, \kappa \, A \,\left(4\, g\, e^{W}\, X\, \sin (4 \theta)+\sqrt{2}\, \kappa\,  (\cos (4 \theta )-3)\right)\\
&-3\, \sqrt{2}\, \kappa ^2\, A^2\, (\cos (4 \theta)-3)+ 8\, g\, e^{W}\,   X \,\left(2 \,e^{W}\, f\, (3\, \cos (4 \theta )-1)-3\, \kappa \, \sin (4 \theta )\right)]\biggr]\,,\\
W'\,=\,& \frac{ e^{V-2 W}}{40\, g\, X\, \cos (2 \theta)}\, \biggl[-2\, \kappa \, A\, \left(4 \,g\, e^{W} X\, \sin (4 \theta )+\sqrt{2}\, \kappa \, (\cos (4 \theta)-8)\right)\\
&+\sqrt{2}\, \kappa ^2 \,A^2 \,(\cos (4 \theta)-8)-8\, g\, e^{W} \,X \, \left(2 e^{W} \,f\, (\cos (4 \theta)-2)-\kappa \,  \sin (4 \theta)\right)\biggr]\,,\\
Y'\,=\,&\frac{ e^{V-2 W}\,Y}{160\, g\, X\,\cos (2 \theta )}\,\biggl[ 6\, \kappa \, A \,\left(4\, g\, e^{W}\, X\, \sin (4 \theta)+\sqrt{2}\, \kappa\,  (\cos (4 \theta )-3)\right)\\
&-3\, \sqrt{2}\, \kappa ^2\, A^2\, (\cos (4 \theta)-3)+ 8\, g\, e^{W}\,   X \,\left(2 \,e^{W}\, f\, (3\, \cos (4 \theta )-1)-3\, \kappa \, \sin (4 \theta )\right)]\biggr]\,,\\
\theta'\,=\,&\frac{ e^{V-2 W}}{80\, g\, X}\, \biggl[-30\, \kappa \, A \left(\sqrt{2}\, \kappa \, \sin (2 \theta )-4\, g \,e^{W} \,X \,\cos (2 \theta )\right)+15\, \sqrt{2}\, \kappa ^2\, A^2 \,\sin (2 \theta)\\
&-8\, g\, e^{W}\, X\,   \left(4 \,e^{W} \,X\, \sin (2 \theta )\, D_X\,f+26\, e^{W} \,f\, \sin (2 \theta)+15\, \kappa \, \cos (2 \theta)\right)\biggr]\,,\\
k'\,=\,& -\frac{3 \,e^{3 U+V-2 W}}{2\, g\, X^3}\,\biggl[6\, \kappa \, A \,\left(\sqrt{2} \,\kappa \, \tan (2 \theta)-2 \, g\, e^{W}\, X\right)\\
&-3 \, \sqrt{2}\, \kappa ^2\, A^2\, \tan (2 \theta )+4\, g\, e^{W}\, X\, \left(4\, e^{W} \,f\,
   \tan (2 \theta)+3 \,\kappa \,\right)\biggr]\,,\\
l'\,=\,&\frac{24 \, \sin (2 \theta ) \,e^{V+3 W} }{5\, X^2}\,\left(f\,-\,X\, D_X\,f\right)\,,\\
X'\,=\,&\biggl[- 6\, \kappa \, A \, \left(4 \, g \, e^{W} \,X \, \sin (4 \theta )+\sqrt{2} \, \kappa \, (\cos (4 \theta )-3)\right)
+3 \, \sqrt{2} \,\kappa ^2 \,A^2\, (\cos (4 \theta )-3)\\
&-8\, g\, e^{W}\,  X\, \left(4 \,e^{W} \,X\, \cos ^2(2 \theta) \,D_X\,f-8\, e^{W}\, f\, \sin ^2(2 \theta )-3\, \kappa \, \sin (4 \theta)\right)\biggr]\,\frac{ e^{V-2 W}}{80\, g\,\cos (2 \theta )} \,,\\
A'\,=\,&-\frac{ e^{V- W}}{10\,\sqrt{2}\,\kappa}\, \biggl[10\, \kappa \, A\, \left(\sqrt{2}\, \kappa \, \sin (2 \theta )-4\, g\, e^{W}\, X\, \cos (2 \theta)\right)-5\, \sqrt{2} \,\kappa ^2 \,A^2\, \sin (2 \theta)\\
&+8 \,g\, e^{W}\, X\,  \left(2\, e^{W} \,X\, \sin (2 \theta) \,D_X\,f+8\, e^{W}\, f\, \sin (2 \theta )+5 \,\kappa \, \cos (2 \theta )\right) \biggr]\,.
\end{split}
\label{flow:mkw21vector}
\ee

\newpage
\section{Flow equations in presence of vectors: $\mathrm{AdS}_{3}\times S^3$}
\label{floweq:AdS3_S3_vector}

In this appendix we present the first-order flow equations that have been solved numerically in section \ref{AdS3vector}.
Given the \emph{Ansatz} \eqref{ansatz:AdS3vector} with the usual Killing spinor of the form \eqref{Kspinor} and satisfying \eqref{gamma3proj} and \eqref{vectorproj}, the system of first-order flow equations is given by
\be
\begin{split}
 U'\,=\,&\frac{ e^{V-2 W}}{80\, g\, X\,\cos (2 \theta )}\,\biggl[ 6\, \kappa \, A \,\left(4\, g\, e^{W}\, X\, \sin (4 \theta)+\sqrt{2}\, \kappa\,  (\cos (4 \theta )-3)\right)-3\, \sqrt{2}\, \kappa ^2\, A^2\, (\cos (4 \theta)-3)\\
&+ 8\, g\, e^{W}\,   X \,\left(2 \,e^{W}\, f\, (3\, \cos (4 \theta )-1)-3\, \kappa \, \sin (4 \theta )+4\,e^{W-U}\,L\,\sin(2\theta)\right)\biggr]\,,\\
W'\,=\,& \frac{ e^{V-2 W}}{40\, g\, X\, \cos (2 \theta)}\, \biggl[-2\, \kappa \, A\, \left(4 \,g\, e^{W} X\, \sin (4 \theta )+\sqrt{2}\, \kappa \, (\cos (4 \theta)-8)\right)+\sqrt{2}\, \kappa ^2 \,A^2 \,(\cos (4 \theta)-8)\\
&-8\, g\, e^{W} \,X \, \left(2 e^{W} \,f\, (\cos (4 \theta)-2)-\kappa \,  \sin (4 \theta)+3\,e^{W-U}\,L\,\sin(2\theta)\right)\biggr]\,,\\
Y'\,=\,&\frac{ e^{V-2 W}\,Y}{160\, g\, X\,\cos (2 \theta )}\,\biggl[ 6\, \kappa \, A \,\left(4\, g\, e^{W}\, X\, \sin (4 \theta)+\sqrt{2}\, \kappa\,  (\cos (4 \theta )-3)\right)-3\, \sqrt{2}\, \kappa ^2\, A^2\, (\cos (4 \theta)-3)\\
&+ 8\, g\, e^{W}\,   X \,\left(2 \,e^{W}\, f\, (3\, \cos (4 \theta )-1)-3\, \kappa \, \sin (4 \theta )+4\,e^{W-U}\,L\,\sin(2\theta)\right)\biggr]\,,\\
\theta'\,=\,&\frac{ e^{V-2 W}}{80\, g\, X}\, \biggl[-30\, \kappa \, A \left(\sqrt{2}\, \kappa \, \sin (2 \theta )-4\, g \,e^{W} \,X \,\cos (2 \theta )\right)+15\, \sqrt{2}\, \kappa ^2\, A^2 \,\sin (2 \theta)\\
&-8\, g\, e^{W}\, X\,   \left(4 \,e^{W} \,X\, \sin (2 \theta )\, D_X\,f+26\, e^{W} \,f\, \sin (2 \theta)+15\, \kappa \, \cos (2 \theta)-15\, L\, e^{W-U}\right)\biggr]\,,\\
k'\,=\,& -\frac{3 \,e^{3 U+V-2 W}}{2\, g\, X^3}\,\biggl[6\, \kappa \, A \,\left(\sqrt{2} \,\kappa \, \tan (2 \theta)-2 \, g\, e^{W}\, X\right)\\
&-3 \, \sqrt{2}\, \kappa ^2\, A^2\, \tan (2 \theta )+4\, g\, e^{W}\, X\, \left(4\, e^{W} \,f\,
   \tan (2 \theta)+3 \,\kappa \,-2\, L\, e^{W-U} \,\sec (2 \theta )\right)\biggr]\,,\\
l'\,=\,&-\frac{6  \,e^{V+3 W} }{5\, X^2}\,\biggl[5\,L\,e^{-U}-4\, \sin (2 \theta )\,\left(f\,-\,X\, D_X\,f\right)\biggr]\,,\\
X'\,=\,&\frac{ e^{V-2 W}}{80\, g\,\cos (2 \theta )}\,\biggl[- 6\, \kappa \, A \, \left(4 \, g \, e^{W} \,X \, \sin (4 \theta )+\sqrt{2} \, \kappa \, (\cos (4 \theta )-3)\right)+3 \, \sqrt{2} \,\kappa ^2 \,A^2\, (\cos (4 \theta )-3)\\
&-8\, g\, e^{W}\,  X\, \left(4 \,e^{W} \,X\, \cos ^2(2 \theta) \,D_X\,f-8\, e^{W}\, f\, \sin ^2(2 \theta )-3\, \kappa \, \sin (4 \theta)+4\, L\, e^{W-U}\, \sin (2 \theta)\right)\biggr] \,,\\
A'\,=\,&-\frac{ e^{V- W}}{10\,\sqrt{2}\,\kappa}\, \biggl[10\, \kappa \, A\, \left(\sqrt{2}\, \kappa \, \sin (2 \theta )-4\, g\, e^{W}\, X\, \cos (2 \theta)\right)-5\, \sqrt{2} \,\kappa ^2 \,A^2\, \sin (2 \theta)\\
&+8 \,g\, e^{W}\, X\,  \left(2\, e^{W} \,X\, \sin (2 \theta) \,D_X\,f+8\, e^{W}\, f\, \sin (2 \theta )+5 \,\kappa \, \cos (2 \theta )-5 L e^{W-U}\right) \biggr]\,.
\end{split}
\label{flow:AdS3vector}
\ee

\newpage

%%%%%%%%%%%%%%%%%%%%%%%%%%%%%%%%%%%%
%
% Bibliography
%
%%%%%%%%%%%%%%%%%%%%%%%%%%%%%%%%%%%%

\small

\bibliography{references}
\bibliographystyle{utphys}

\end{document}